\documentclass[fleqn,usenatbib]{mnras}

\usepackage{newtxtext,newtxmath}

\usepackage[T1]{fontenc}
\usepackage{ae,aecompl}

\usepackage{graphicx}	
\usepackage{amsmath}	
\usepackage{amssymb}	
\usepackage{makecell}
\usepackage{booktabs}
\usepackage{tabularx}
\usepackage{xfrac}
\usepackage[dvipsnames]{xcolor}
\usepackage{supertabular}

\usepackage{tikz}
\usetikzlibrary{matrix,chains,positioning,decorations.pathreplacing,arrows}

\newcolumntype{Y}{>{\centering\arraybackslash}X}
\let\vec\mathbf

\title[Machine learning identification of submillimetre galaxies]{A machine learning approach for identifying the counterparts of submillimetre galaxies and applications to the GOODS-North field}
\author[R. H. Liu et al.]{
Ruihan Henry Liu,$^{1}$\thanks{E-mail: rhliu@phas.ubc.ca}
Ryley Hill,$^{1}$
Douglas Scott,$^{1}$
Omar Almaini,$^{2}$
Fangxia An,$^{3,\, 4}$\newauthor
Chris Gubbels,$^{1}$
Li-Ting Hsu,$^{5}$
Lihwai Lin,$^{5}$
Ian Smail,$^{4}$
Stuart Stach$^{4}$
\\
$^{1}$Department of Physics and Astronomy, University of British Columbia, 6224 Agricultural Road, Vancouver, V6T 1Z1, Canada\\
$^{2}$School of Physics and Astronomy, University of Nottingham, University Park, Nottingham, NG7 2RD, UK\\
$^{3}$Department of Physics and Astronomy, University of the Western Cape, Robert Sobukwe Road, 7535 Bellville, Cape Town, South Africa\\
$^{4}$Centre for Extragalactic Astronomy, Department of Physics, Durham University, South Road, Durham, DH1 3LE, UK\\
$^{5}$Academia Sinica, Institute of Astronomy \& Astrophysics (ASIAA), P.O.~Box 23-141, Taipei, 10617, Taiwan\\
}

\date{Accepted XXX. Received YYY; in original form ZZZ}

\pubyear{2018}

\begin{document}
\label{firstpage}
\pagerange{\pageref{firstpage}--\pageref{lastpage}}
\maketitle

\begin{abstract}
Identifying the counterparts of submillimetre (submm) galaxies (SMGs) in multiwavelength images is a critical step towards building accurate models of the evolution of strongly star-forming galaxies in the early Universe. However, obtaining a statistically significant sample of robust associations is very challenging due to the poor angular resolution of single-dish submm facilities. Recently, a large sample of single-dish-detected SMGs in the UKIDSS UDS field, a subset of the SCUBA-2 Cosmology Legacy Survey (S2CLS), was followed up with the Atacama Large Millimeter/submillimeter Array (ALMA), which has provided the resolution necessary for identification in optical and near-infrared images. We use this ALMA sample to develop a training set suitable for machine-learning (ML) algorithms to determine how to identify SMG counterparts in multiwavelength images, using a combination of magnitudes and other derived features. We test several ML algorithms and find that a deep neural network performs the best, accurately identifying 85\,per cent of the ALMA-detected optical SMG counterparts in our cross-validation tests.  When we carefully tune traditional colour-cut methods, we find that the improvement in using machine learning is modest (about 5\,per cent), but importantly it comes at little additional computational cost.  We apply our trained neural network to the GOODS-North field, which also has single-dish submm observations from the S2CLS and deep multiwavelength data but little high-resolution interferometric submm imaging, and we find that we are able to classify SMG counterparts for 36/67 of the single-dish submm sources. We discuss future improvements to our ML approach, including combining ML with spectral energy distribution-fitting techniques and using longer wavelength data as additional features. 

\end{abstract}

\begin{keywords}
methods: data analysis -- galaxies: starburst -- submillimetre: galaxies
\end{keywords}


\section{Introduction}
\label{sec:Intro}

\vspace{0.5cm}

The submillimetre (submm) window has become an important waveband for extragalatic astronomy due to the discovery of bright submm galaxies (SMGs). These galaxies appear to be among the earliest and most actively star-forming galaxies in the Universe, often reaching luminosities of more than 10$^{13}$\,L$_{\odot}$ (over 100 times that of our Milky Way galaxy) and star-formation rates (SFRs) greater than a few hundred M$_{\odot}$\,yr$^{-1}$ \citep[e.g.,][]{blain2002,magnelli2012,swinbank2014,mackenzie2017,michalowski2017} around redshifts 2--3, corresponding to only a few billion years after the Big Bang \citep[e.g.,][]{chapman2005,simpson2014,simpson2017}.

Observations of these SMGs are most easily made using single-dish submm telescopes, such as the continuum imaging instruments SCUBA-2 \citep{holland2013} on the James Clerk Maxwell Telescope (JCMT), and the Large Apex BOlometer CAmera \cite[LABOCA;][]{sirigno2009} on the Atacama Pathfinder EXperiment. However, such single-dish surveys are usually low in angular resolution, typically around 15--20\,arcsec; finding multi-wavelength counterparts to these sources is therefore difficult, since there could dozens of optically-detected galaxies within the submm beamsize. 

This identification problem was first tackled using observations with interferometres at radio wavelengths \citep[e.g.,][]{chapman2001,ivison2002,chapman2002,chapman2003,bertoldi2007,smail2000,lindner2011}. In the radio, synchrotron emission is linked to supernovae, which is correlated with far-infrared (FIR) emission from dust \citep[e.g.,][]{condon1992,yun2001,ivison2010, magnelli2015}. While these radio-derived studies paved the way for progress in multiwavelength detection, probabilistic arguments were still required, since the submm emission could not be directly resolved.

In order to achieve the arcsecond and sub-arcsecond resolution required for directly detecting individual SMGs, one must turn to submm interferometres such as the the Submillimeter Array \citep[SMA;][]{ho2004} and the Atacama Large Millimeter/submillimeter Array \citep[ALMA;][]{wootten2009}. Unfortunately, locating statistically significant numbers of SMGs over large areas of sky with such telescopes is prohibitive in both time and resources. A more efficient way of utilizing these higher resolution instruments is therefore to follow up bright individual SMGs previously found in single-dish surveys \citep{barger2012,smolcic2012,hodge2013,simpson2015,hill2018,Stach2018}. With such data, one can gather samples over square-degree scales and accurately pinpoint many hundreds of early-Universe SMGs. Nevertheless, there are much larger samples of SMGs than can be efficiently followed up one-by-one using interferometres.

Since we already have detailed information about the counterparts to SMGs in samples with interferometric data, we can use that information to help find the correct identifications in surveys lacking such high-resolution imaging.  In other words, we can use known counterparts as a training set for identifying the SMGs in single-dish surveys, through the application of deep-learning techniques.

Machine learning (ML) has seen growing interest in astronomy over the past decade or so.  The rapid increase in the size of astronomical data sets has led to a need for fast, automated algorithms to extract relevant information, and astronomers are increasingly turning to ML techniques to achieve their goals.  Examples include 
finding structures in galaxy surveys and simulations \citep[e.g.,][]{barrow1985,hajian2015}, identifying cosmic ray artefacts in images \citep[e.g.][]{salzberg1995},
detection of sources in $\gamma$-ray data \citep[e.g.][]{campana2008}, classification of stellar spectra \citep[e.g.][]{bailerjones1997}, fitting photometric redshifts of galaxies \citep[e.g.][]{collister2004}, producing mock galaxy catalogues \cite[e.g.][]{xu2013}, approximating star-formation histories \citep[e.g.][]{cohn2018} and even use as a proxy for simulations of galaxy formation \cite[e.g.][]{kamdar2016}.

An application closely related to the current paper is source detection and classification \citep[e.g.,][]{sextractor,andreon2000}, which already has a long history \citep[see e.g.,][]{odewahn1992,storrielombardi1992}. Automated star-finders have been developed using decision trees \citep{ball2006}, and more recently the use of algorithms such as support-vector machines (SVMs) and convolutional neural networks (CNNs) have been used to find sources down to even fainter magnitudes \citep{krakowski2016,kim2017}. 

A similar task involves classifying images of galaxies based on their morphologies (e.g. spiral, elliptical, irregular). This problem was studied in detail in astronomy using early neural networks that contained only a few layers \citep{naim1995,odewahn2002,calleja2004}, and it was initially found to be no more accurate than traditional weighted regression schemes. Later, when SVMs appeared on the scene, there was a significant improvement over the rudimentary neural networks used in the past \citep{huertas2008,huertas2009}, but it was not until CNNs were trained to distinguish different galaxies from one another \citep{dieleman2015,huertas2015,sanchez2018} that ML became an effective and convincing tool for image recognition in galaxy surveys.

Recently, \citet{an2018} used interferometric data from an ALMA survey of the UKIDSS UDS (hereafter UDS) field \citep{Stach2018}, designed to follow up sources detected in the SCUBA-2 Cosmology Legacy Survey \citep[S2CLS;][]{geach2016}, to train an ML algorithm to find the optical and near-infrared (NIR) counterparts to single dish-detected SMGs. These authors followed up 716 SMGs detected in the UDS field as part of the SCUBA-2 Cosmology Legacy Survey \citep[S2CLS;][]{geach2016}, and matched each galaxy to overlapping optical and NIR images. They then used an SVM to separate the known SMGs from the non-SMGs that happened to lie within the single-dish emission region. It was found that the SVM was able to achieve $77.2\pm 4.7$ and $82.0\pm 4.9$ per cent in precision and recall, respectively, where precision and recall are two evaluation metrics for binary classifiers (described in Section~\ref{sec:Validation}). Other supplementary inputs, such as radio detections, were required to increase the performance.

To further develop this approach, in this paper we will test a variety of different ML algorithms
to determine the best method for classifying SMGs. In Section~\ref{sec:Trainset} we discuss how we use existing data to develop a training set, in Section~\ref{sec:ML} we explore various ML algorithms, data-augmentation techniques and validity tests, and in Section~\ref{sec:Results} we evaluate how well these ML algorithms perform. Then, in Section~\ref{sec:NewPred}, we apply our fully trained deep neural network to the GOODS-North (hereafter GOODS-N) field, which contains a similar catalogue of single-dish SMGs from the S2CLS and multiwavelength images. In Section \ref{sec:future} we discuss possible improvements to our ML approach to SMG classification, and we conclude in Section~\ref{sec:Conclusions}.

\section{Developing the training and prediction data sets}
\label{sec:Trainset}

\subsection{Training catalogues}

\subsubsection{The single-dish survey}

The first component of our ML training set is data from a large single-dish submm survey. Following the previous work carried out by \citet{an2018}, we used the subset of the S2CLS covering the UDS field, which encompasses an area of 0.9\,deg$^2$ at 850\,$\mu$m, in which there are 296,007 catalogued optical galaxies. The resolution of the SCUBA-2 instrument is 14.8\,arcsec, meaning that a typical blank-field optical image will find between five and ten galaxies within a single SCUBA-2 beam. In total, 1,084 submm sources were detected in the UDS field by SCUBA-2 with signal-to-noise ratio ${\rm S/N}\,{>}\,3.5$, and with flux densities ranging from 2 to 17\,mJy (not including a 50\,mJy strong gravitational lens, which we did not incorporate into this study).

\subsubsection{Interferometric follow-up}

All ${\rm S/N}\,{>}\,4.0$ S2CLS sources in the UDS field (716 in all) have been followed up with ALMA at 870\,$\mu$m in the AS2UDS survey \citep{Stach2018}. The maps produced by these observations had a tapered angular resolution of 0.5\,arcsec, more than sufficient to match that of typical optical and NIR maps, which is needed to directly connect SMGs to counterparts in optical and NIR catalogues. In the 716 ALMA follow-up maps, 608 contained ALMA-detected SMGs, with some images containing more than one. In all, a total of 695 SMGs were detected above $4.3\sigma$ by ALMA. 

\subsubsection{The multiwavelength follow-up}
\label{sec:multiwavelength_follow_up}

The optical and NIR data for the UDS field comes from a variety of surveys carried out by several telescopes over the past decade. Briefly, the United Kingdom Infrared Telescope 
has covered the field in the $J$ (1220\,nm), $H$ (1630\,nm) and $K$ (2190\,nm) bands from the UKIDSS survey \citep{lawrence2007} with 3$\sigma$ depths of 26.2, 25.7 and 25.9, respectively; the Visible and Infrared Survey Telescope for Astronomy 
has provided data in the $Y$ band (1020\,nm) at a 3$\sigma$ depth of 25.3; the $u$ (360\,nm) band has been covered by the Canada-France-Hawaii Telescope telescope down to a 3$\sigma$ depth of 27.3, 
 and from the Subaru telescope we use observations in the $B$ (445\,nm), $V$ (551\,nm), $R$ (658\,nm), $i$ (806\,nm) and $z$ (900\,nm) bands, where the 3$\sigma$ depths are 28.4, 27.8, 27.7, 27.7 and 26.6, respectively. 
Finally, a catalogue of NIR sources within the UDS has been obtained from the {\it Spitzer\/} mission at 3.6\,$\mu$m (down to 24.8, 3$\sigma$) and 4.5\,$\mu$m (down to 24.6, 3$\sigma$). 
The details of these observations will be discussed in \textcolor{red}{Almaini et al.~(in prep.)}. Using this UDS catalogue, \citet{an2018} were able to match 514 of the 698 ALMA-detected SMGs from the AS2UDS survey to their multiwavelength counterparts, spanning the 12 wavebands described above (the details of the photometric-matched catalogue will be described in \textcolor{red}{Hartley et al.~in prep.}). We used this AS2UDS catalogue to make the training set in our paper.

To create the non-SMG portion of the training set, we took all of the sources not identified as SMGs by ALMA from the multiwavelength catalogue that were within a 7.0\,arcsec radius of the 608 SCUBA-2 sources with ALMA detections. We chose this radius as it is gave a search region with a diameter equal to the effective full-width at half-maximum (FWHM) of the SCUBA-2 beam, and the probability of finding the correct position of an SMG with a $\rm{S/N}\,{>}4.3$ that lies within this area is 99.6 per cent \citep[][]{Stach2018}. This is very similar to the approach taken by \citet{an2018}, except they used a slightly larger search radius of 8.7\,arcsec, which is the radius of the ALMA primary beam at 870\,$\mu$m. We omitted SCUBA-2 sources that lacked an ALMA match, since the multiwavelength sources in those regions could still correspond to a weaker than $4.3\sigma$ SMG. The UDS catalogue has a built-in star classification algorithm from \texttt{SExtractor}, which classifies the likelihood of sources being stars versus galaxies based on $K$-band photometry. We used this built-in tool to filter out sources with a >50 per cent likelihood of being a star. The resulting training set for our ML testing contained a total of 1439 sources: 514 SMGs; and 925 non-SMGs, which are defined as multiwavelength sources within the 7\,arcsec radius SCUBA-2 beams but without >4.3$\sigma$ ALMA detection.


\subsection{Prediction catalogue}

\label{sec:PredGOODSN}


Once an ML algorithm has been trained to identify SMG counterparts, we can use it on fields where no submm interferometric follow-up data exist, but optical and NIR catalogues are available. One such field is the Great Observatories Origins Deep Survey North (GOODS-N) field. The GOODS-N field is one of the S2CLS regions, and therefore has the necessary single-dish submm observations. There is also a catalogue of optical- and NIR-detected galaxies detected in this field using a combination of ground-based imaging from the Subaru and the Canada-France-Hawaii telescopes, and space-based imaging from {\it Spitzer\/} (see \citealt{hsu2019} for details), allowing us to use the best fully trained algorithm to locate the multiwavelength counterparts of submm sources in the GOODS-N field. For reference, the 3$\sigma$ depths obtained in this catalogue are 27.6 in the $U$ band, 27.6 in the $B$ band, 27.0 in the $V$ band, 27.6 in the $R$ band, 26.4 in the $i$ band, 26.0 in the $z$ band, 26.2 in the $Y$ band, 25.3 in the $J$ band, 24.8 in the $H$ band, 25.0 in the $K$ band, 25.6 in the 3.6\,$\mu$m band and 25.6 in the 4.5\,$\mu$m band. In total, this field contains 67 SCUBA-2-detected submm sources with flux densities ranging from 3 to 13\,mJy. This catalogue contains a total of 399 multiwavelength sources within these 67 beams. 

The depths obtained by the ground-based observations in the UDS field are a factor of about 2 deeper than in the GOODS-N field, and conversely the {\it Spitzer\/} imaging in the GOODS-N filed is a factor of about 2 deeper than in the UDS field. Although it would be ideal to use training and prediction sets of exactly equal depth, we found that such a limit would drastically reduce the size of both our UDS training set and our GOODS-North prediction set. However, we did check to see if reducing the size of the training and testing data sets to make their depths equal had any appreciable affects on our results, and we found that indeed there were none. We therefore chose to use the full scope of the available data; we will further discuss our tests in Section \ref{sec: GOODSN_SMGs}.


In addition to the multiwavelength data, the GOODS-N field has seen a partial interferometric follow-up from the SMA \citep{cowie2017}. We will utilize this interferometric catalogue to further verify our results from the ML approach, and provide a ML classified catalogue of the multiwavelength sources.


\subsection{Feature selection}

It is well known that any ML algorithm can only be as good as the features that it uses to learn. 
Thus, while training, it is important to strike a balance between using lots of features, most of which are somewhat useful, and using far too many useless features that just become noise; it is also important not to compress large numbers of features together or to re-use features.

\begin{figure*}
  \includegraphics[width=16cm]{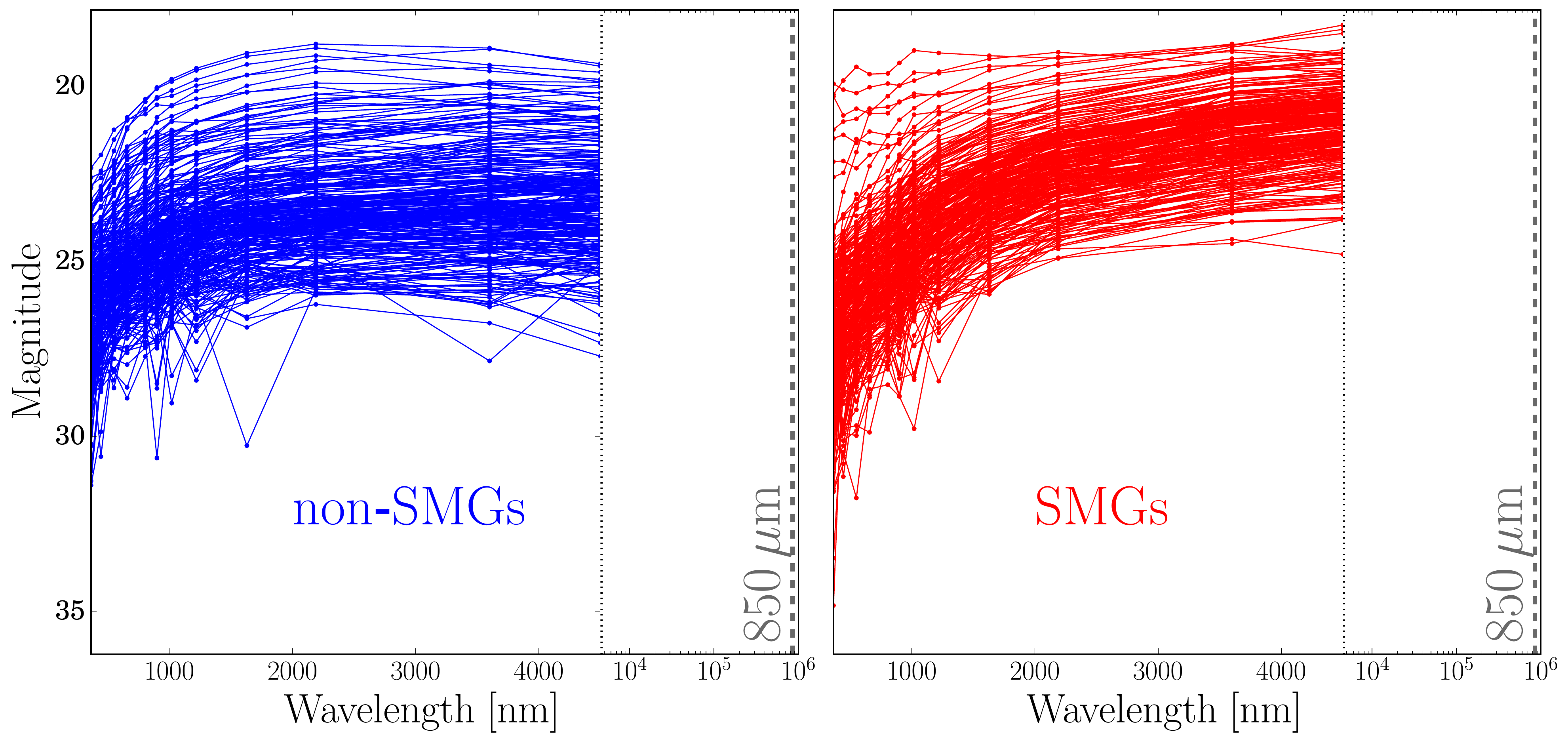}
  \caption{Plot of the magnitudes of our galaxies in the training set as a function of wavelength. The left panel shows the non-SMGs, and the right panel shows the SMGs. It is clear that the SMGs are generally brighter at longer wavelengths (i.e. redder), which is largely the characteristic that our ML algorithm will use (along with detailed colour information corresponding to shapes and breaks in different redshift ranges) to predict which galaxies will be bright at 850\,$\mu$m (indicated by the dashed black line).}
  \label{fig:sed}
\end{figure*}

Here we used as features the 12 magnitude measurements from \textcolor{red}{Hartley et al.~(in prep.)}, as well as the parent SCUBA-2 flux-density measurements. These are clearly the most useful features to use, since we are effectively training a model to predict which galaxies are bright at 850\,$\mu$m based on their brightness at optical and NIR wavelengths; this is similar to fitting optical/NIR spectral energy distributions (SEDs) and extrapolating to the submm. Any additional information about the galaxies (photometric redshifts, luminosities, etc.) will come from assuming some sort of SED, which will hence be reusing the same information. In Fig.~\ref{fig:sed} we show `un-extrapolated' SEDs for our galaxies (i.e. the magnitudes in the observed frame), colour-coded by SMG/non-SMG classification, to demonstrate this idea. ML algorithms will take into account all the available features and put heavier weights on the features with more prediction power, while still retaining less important features in a lower capacity. Here we are assuming that there are no issues with noise in the data, a topic that will be further discussed in Section \ref{sec:future}. In a sense, these features will be used by the ML algorithms to estimate photometric redshifts and specific rest-frame colours that predict dust-enshrouded star formation. We note that in the previous study of SMG identification with ML by \citet{an2018}, they chose touse five features: photometric redshifts (derived from the above 12 optical/NIR magnitudes); $H$-band magnitudes; $J-K$ colours; $K-$[3.6\,$\mu$m] colours; and [3.6\,$\mu$m]$-$[4.5\,$\mu$m] colours.

In addition to the flux features, we also want to include the angular separations of the optical/NIR sources from the positions of the SCUBA-2 centroids (i.e.~the position of the brightest pixel within the 14\,arcsec area). We expect the positional accuracy to be approximately $\sigma_{\rm pos}\,{=}\,0.6$ FWHM/(S/N) in each coordinate, and hence the probability distribution for correct IDs should follow $P(r) \propto  r\times \exp(-r^2/2\sigma_{\rm pos}^2)$ \citep{ivison2007}. In these equations FWHM refers to the beamsize of SCUBA-2 at 850\,$\mu$m and S/N is the signal-to-noise ratio of the SCUBA-2 source detection. Although some single-dish sources resolve into multiple galaxies, rendering implementation of the above equation more complicated, we believe that the inclusion of such sources will be beneficial overall. Indeed, by plotting the distributions of the angular separations for SMGs and non-SMGs, we found that the best way to incorporate this information was by including as a feature the angular separation scaled by $\sigma_{\rm pos}$. In addition, we compared a training set that used $\sigma_{\rm pos}$ as a feature to one that did not, using the algorithms discussed in Section \ref{sec:Algorithms}, and we found that the inclusion of this feature increased the precision and recall (see Section \ref{sec:Validation} for definitions of these metrics) by around 1--3\,per cent.

In the end, our final training/predicting data sets are comprised of 14 features: magnitudes in the $U$, $B$, $V$, $R$, $i$, $z$, $Y$, $J$, $H$, $K$, 3.6\,$\mu$m, and 4.5\,$\mu$m bands, the 850\,$\mu$m flux densities, and the scaled angular separations.

Unfortunately, a number of the galaxies in these data sets were too faint to be detected in all 14 bands, resulting in examples with incomplete features. This is a common issue in data science, and in Section~\ref{sec:Impute} we will explore methods to cope with this issue. In our training set of 1,483 sources, 301/513 SMGs and 856/970 non-SMGs have complete data across all fourteen features; similarly, in our GOODS-N prediction set, 290/399 sources have complete photometric data.

In order to maximize the number of unique samples used for training, we chose to use all of the fully-featured sources for training, without making any S/N cuts; this means that some of our training examples have relatively low S/N, down to about $3\sigma$. The motivation behind this choice is that there is still important information present in the photometric data when averaged over a large number of sources. To ensure that the low S/N data are not fitting incorrect models or causing overfitting issues, in Section \ref{sec:CrossVal_results} we verify our results with validation tests to ensure that the classifiers will still maintain performance on high S/N data. In Section \ref{sec:missingfeatures} we also discuss future improvements in which photometric uncertainties can be accounted for in ML algorithms.


\subsection{Feature analysis}
\label{sec:Data Analysis}

\begin{figure*}
  \includegraphics[width=16cm]{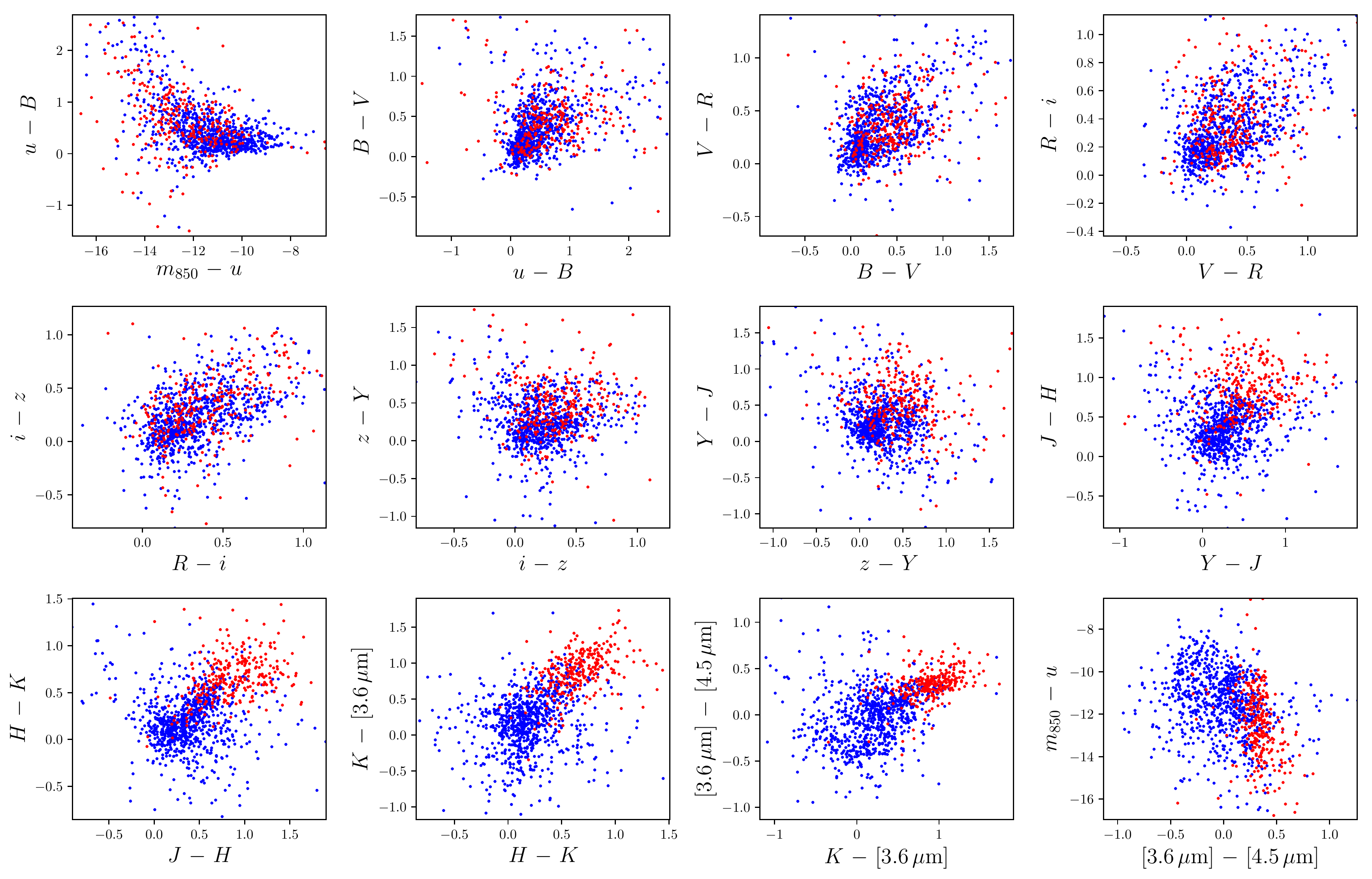}
  \caption{Plot of the data for the training set in a variety of 2D colour spaces. Red points represent SMG sources, while blue points represent non-SMGs. The distributions of the features show a clear divide between SMGs and non-SMGs, especially among the redder colours. We could already effectively separate red from blue by applying a few colour cuts by hand. The question is how much better ML algorithms will perform compared with traditional separation methods. We note that the finite depths of the observations lead to unavoidable selection biases (i.e.~our data do not contain colours redder or bluer than certain limits set be telescope sensitivities); these biases should in principle apply equally to both SMG and non-SMGs, and as such ML algorithms can still distinguish between them.}
  \label{fig:colorspacedist}
\end{figure*}

Before starting our ML training, it is useful to determine how much separation is visible in the marginal distributions, such that we have some idea of a lower bound on our expected classification accuracy, in order to confirm that an algorithm can correctly separate the data.
There are of course a very large number of ways to slice a 14-dimensional parameter space; however, in practice the galaxy types will be distinguished through specific slopes and breaks in their rest-frame SEDs. A simple and easy-to-interpret way to highlight this is to plot the data in a series of 2D colour spaces, as shown in Fig.~\ref{fig:colorspacedist}. 

We can see that a separation between SMG and non-SMG sources does clearly exist, especially in the redder colours and wavelengths (as expected since we are looking for galaxies that are brighter at longer wavelengths, as well as typically being dustier). We can therefore be confident that ML algorithms will be capable of separating the sources with a reasonable degree of accuracy. The traditional astronomical approach would be to apply colour cuts to separate the classes of sources. ML should be able to use more of the available information (including weak separation in multiple dimensions that would not be visible to the naked eye) to separate the SMGs and non-SMGs. What we want to investigate is how much better ML can perform. In Section~\ref{sec:Results} we will show the benefits of using all features, compared to limiting ourselves to just the obvious ones.	

\section{Machine-learning algorithms, optimization, and validation}
\label{sec:ML}

\subsection{Algorithms}
\label{sec:Algorithms}

Two ML algorithms have already been tested with SMG classification. \citet{an2018} explored the effectiveness of an SVM and the more recently developed boosted decision-tree (BDT) algorithm called \texttt{XGBoost}\footnote{\url{https://github.com/dmlc/xgboost}} \citep{friedman2001}. The SVM model creates hyperplanes in the $n$-dimensional feature space to separate the input samples into different groups. \texttt{XGBoost,} uses an ensemble of decision trees to score each sample and decide which class it belongs to, using gradient boosting to optimize the loss function (which measures how predictive the model is). An obvious advantage of the latter algorithm is the fact that it can take into account samples with missing features (which indeed is an issue with our training set).


In this paper, we will expand on these models by incorporating a number of other ML algorithms, along with the expanded set of features. In addition to the SVM and BDT algorithms, we will include logistic regression (LR), Gaussian-naive Bayes (GNB), $k$-nearest neighbors (KNN), regular decision trees (DT), random forests (RF), linear discriminant analysis (LDA), and finally, a few neural networks (NN) with different structures. We give further details regarding the traditional ML models in Appendix~\ref{app:ML_models}, while our deep-learning methodology and neural-network architecture is described in Appendix~\ref{app:DNN}. By testing this many classification methods, we expect to be able to pick the best algorithm for our particular application. It is worth noting that some caution is required here, since we are testing so many different ML algorithms; overfitting, a phenomena where a classification algorithm picks up on random patterns in order to describe the training set with near 100 per cent accuracy \citep[see e.g.,][]{Buduma2015,Chollet2017}, can be an issue in certain scenarios. However, we do not expect overfitting to be an issue here because the ML models that we are testing are relatively simple and do not depend on very many free parameters, as compared to, for instance, a very deep NN with hundreds of thousands of free parameters.

Most of our additional models were implemented using the \texttt{scikit-learn}\footnote{\url{ http://scikit-learn.org}} package for \texttt{Python}, but we constructed our NNs using the \texttt{Keras}\footnote{\url{https://keras.io/}} NN interface library, which is built on top of many of todays popular deep-learning software libraries. We also used Google's \texttt{Tensorflow}\footnote{\url{https://www.tensorflow.org/}} deep-learning library as the \texttt{Keras} deep-learning backend. The modularity of \texttt{Keras} allows the construction of many types and variations of NNs. We have specifically constructed four different NNs of varying sizes and configurations, which are described in more detail in Appendix~\ref{app:DNN}.

\subsection{Hyperparameter optimization}
\label{sec:Hyperparams}

Most ML algorithms contain hyperparameters that are tuned to find the best possible algorithm for a particular type and composition of data. All of the algorithms that we tested (except for the GNB classifier) contain such parameters. Depending on the model, these hyperparameters may include different values for the threshold of classification or different variations in the mathematical models for fittimg the data. In particular, those from \texttt{scikit-learn}, as well as from the \texttt{XGBoost} package, can be tuned simply using a $k$-fold cross-validation method. This procedure divides the training set into $k$ subsets, or folds, and trains on $k-1$ folds while validating on the remaining fold (i.e., similar to bootstrap resampling). This process is repeated $k$ times with each separate set serving as the validation data. The cross-validation procedure is performed on every variation of the model given by different combinations of hyperparameters in order to find the best hyperparameter combination. We chose to set $k=4$ in order to have a more reasonable computation time.  Given the relatively small size of our data set, as well as the relative simplicity of the hyperparameters, we do not split a secondary validation set specifically for hyperparameter optimization, since such a set would decrease the amount of training data, and result in less accurately trained models when assessing model performance. 
After the search, we then chose the set of hyperparameters that achieves the best $F_1$ score (see Section~\ref{sec:Validation} for a definition of $F_1$). 

It is a bit more difficult to tune a NN in comparison to the other, simpler models investigated in this study. Using the \texttt{Keras} package's modular neural-network interface, we found that there are millions of different combinations of hyperparameters that can be changed to affect a NN's performance. Furthermore, large NNs tend to be more prone to overfitting than other simpler models, as their relative complexity means that they are more able to `memorize' the training data rather than learning patterns from it. Therefore, we decided to simplify the process by fixing a few parameters while performing hyperparameter optimization on some others. In order to maximize the effectiveness of the neural-network algorithm, we tested various structures with different numbers of layers, nodes, and regularizations to find those that worked best for our application. In the analysis below, we included the best four such structures alongside our other algorithms. Further discussions on overfitting, as well as details on the exact structures used and neural-network optimization can be found in Appendix~\ref{app:DNN}.

\subsection{Validation}
\label{sec:Validation}

There are many ways to quantitatively assess the performance of a classification algorithm. Here we use four different metrics (treated as percentages) to compare each of the ML algorithms in our tests. We define $n_{\rm correct}$ as the total number of correctly classified samples, $n_{\rm total}$ as the total number of tested samples, $n_{\rm TP}$ is the number of true positives (number of correctly identified SMGs), $n_{\rm FP}$ is the number of false positives (number of non-SMGs incorrectly identified as SMGs), and $n_{\rm FN}$ is the number of false negatives (number of SMGs incorrectly identified as non-SMGs). The totals are such that: $n_{\rm correct}=n_{\rm TP}+n_{\rm TN}$ is the number of correctly identified SMGs plus the number of correctly identified non-SMGs; and $n_{\rm total}=n_{\rm correct}+n_{\rm FP}+n_{\rm FN}$ is the total number of classification attempts. Our metrics are:
 
\begin{enumerate}
\item accuracy, defined as $n_{\rm correct}/n_{\rm total}$; 
\item precision, defined as $n_{\rm TP}/(n_{\rm TP}+n_{\rm FP})$; 
\item recall, defined as $n_{\rm TP}/(n_{\rm TP}+n_{\rm FN})$; 
\item the $F_1$ score, a harmonic mean of the precision and recall, defined as $2 \times (\mathrm{precision}  \times  \mathrm{recall})/(\mathrm{precision}+\mathrm{recall})$.
\end{enumerate} 
\noindent
Since precision and recall are both important metrics for our application, we choose to focus on the $F_1$ score because it combines these two quantities, thus we use it to evaluate the best set of hyperparameters and to choose the best ML algorithm. The $F_1$ score can also take into account training sets where the number of positives may be much smaller than the number of negatives, which is the case here.

To estimate these metrics for each ML algorithm, we use $k$-fold cross-validation with $k=4$. The procedure is identical to the $k$-fold cross-validation used to tune hyperparameters, only the hyperparameters are fixed to their optimal values. We split the labelled data set into four subsets, and train on three, while testing on the remaining one. We then rotate the test set, such that we eventually test on the entire data. In order to ensure consistency in our results, we repeated this process 50 times, thus obtaining a total of 200 estimates of the accuracy, precision, recall and $F_1$ score based on 50 random cuts of the training set. We then calculated the mean and standard deviation of the chains. We note that the standard deviation here is best viewed as a measure of the relative stability of the performance of a model (as measured by the given metric, e.g. $F_1$ score) under varying training/testing sets. This standard deviation should not be interpreted as a Gaussian uncertainty in the true value of the metric for a given model trained on a fixed training set.

In addition to the typical ML validation methods, we also implemented another validation method by testing a trained model on blank fields. Blank fields were created by looking at random points in the UDS field (sufficiently far from known SCUBA-2 sources) and considering all of the multiwavelength galaxies within a SCUBA-2 beam in the same way as for the real sources. The blank-field test should give us a better understanding of the false positive rate of our classifiers, especially when compared to the predictions from an unknown data set (such as the GOODS-N field).

\subsection{Imputation}
\label{sec:Impute} 

The effectiveness of an ML algorithm is somewhat related to the size and completeness of the training set. Models will perform better when given a more complete sample of data to train on. However, many galaxies in our catalogue have missing (i.e.~undetected) magnitudes, and removing these galaxies prior to training might not be ideal because it will decrease the training set size. In order to use our full training set, we can include the samples with missing photometry and figure out a way to fill in these missing features. The process of filling in missing features is called imputation in data science, and there are many different algorithms to accomplish the task. 

Of course, imputed data should not be considered as representations of new real data, and imputation is more effective if missing features are random, while in reality they are due to incomplete coverage, populations of faint sources, or other sample biases; Section \ref{sec:missingfeatures} further discusses the causes of the missing features in our sources. Rather than creating new fully-featured training data, the process of imputation can be thought of as a way to improve the performance of algorithms by increasing the training set size.

In this particular setting, where we are dealing primarily with photometry (rather than other kinds of missing data), another reasonable solution to the problem of non-detections would be to report the non-detections in any case. For example, if a galaxy is detected in the $K$ band but in the $B$ band its flux density is less than the local noise, one could still provide this information as a value and error bar, provided that flux density units are used instead of magnitudes. The fact that there is a number associated with the position of the galaxy in the $B$ band is still useful information, even though the S/N is less than some threshold. In fact, one could further extend the ML approach to use the uncertainties by repeating the analysis a large number of times with realisations drawn from the error distributions, which we discuss in section \ref{sec:missingfeatures}. But here we are working directly from catalogues of sources where non-detections are left blank, so in order to deal with these sources we would have to use imputation to fill in the missing data.

We explored various imputation techniques to see if they improved our metrics in any way. We used the \texttt{fancyimpute} package for \texttt{Python}, which implements a number of different algorithms. These include replacing missing values with the mean or the median, and replacing missing values with the source having the closest mean-square distance (i.e.~$k$-nearest neighbours), as well as complex imputation methods such as Soft Impute \citep{Mazumder2010} and Multivariate Imputation by Chained Equations \citep[MICE;][]{Azur2011}. In order to judge the performance of these techniques, we imputed missing data in the training set only, and calculated their metric scores on non-imputed data. Lastly, we compared the metrics obtained from imputed training sets to the metrics obtained from simply removing sources with missing features, in order to establish the usefulness of imputation with our data.

\subsection{A manual learning algorithm}
\label{sec:Manualclassification}

\begin{figure*}
  \includegraphics[width=2\columnwidth]{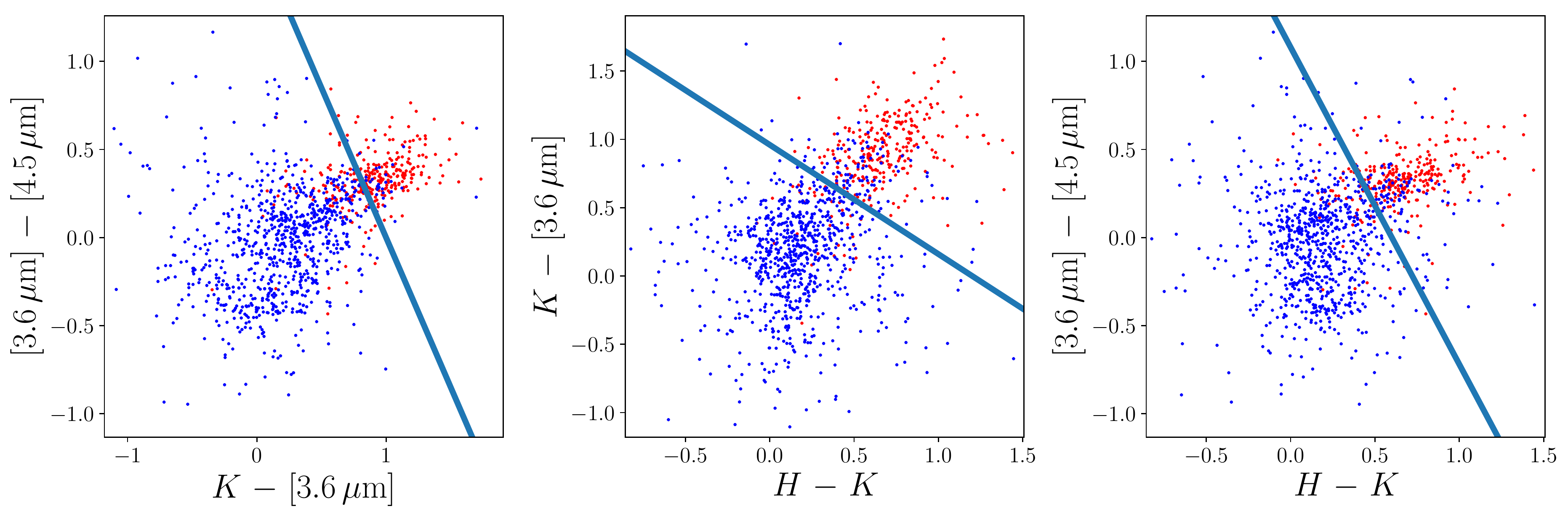}
  \caption{In order to see if `manual learning' can compete with ML, we imposed manual cuts in colour space to separate SMGs from non-SMGs in the training set. The lines shown here have been obtained by searching for the slopes and intercepts that maximize the $F_i$ score. Our manual procedure then classifies a source as an SMG if it satisfies any two of the three linear cuts shown here. The performance of this `manual learning' algorithm is compared to ML algorithms in Section \ref{sec:Results}.}
  \label{fig:manualcut}
\end{figure*}

An important issue to address is whether or not all of the work that goes into training sophisticated ML algorithms (like NNs) is actually worth the effort. In other words, are these algorithms more successful at SMG classification than traditional, manually-identified colour cuts? In the traditional approach, after plotting the data in several colour spaces, it might become apparent that a nearly complete separation exists between SMGs and non-SMGs at redder wavelengths, and that this can be utilised to `learn' how to distinguish SMGs from non-SMGs without any ML at all.

We therefore pitted such a `human' or `manual learning' algorithm against the ML algorithms to see how the performance compares. In Fig.~\ref{fig:manualcut} we show the magnitudes and colours against one another in various combinations, and we looked for spaces where there was an obvious separation between SMGs and non-SMGs. There are three plots where this separation seems clear: $K-$[3.6\,$\mu$m] versus [3.6\,$\mu$m]$-$[4.5\,$\mu$m]; $H-K$ versus $K-$[3.6\,$\mu$m]; and $H-K$ versus [3.6\,$\mu$m]$-$[4.5\,$\mu$m]. Next, we found the slopes and intercepts of lines that best separated the two classes in these three plots by applying the concept of hyperparameter optimization. Iterating over different slopes and intercepts for each of the cuts on the colour space, we found the best set of slopes and intercepts for the three colour cut lines that yielded the best $F_1$ score. We certainly expect ML approaches to do better, since they use information about the separation of sources in all dimensions of the colour space, but the question is whether they perform dramatically better or only slightly better.

\section{Results}
\label{sec:Results}

\subsection{Validation results}
\label{sec:CrossVal_results}

\begin{figure}
  \includegraphics[width=\columnwidth]{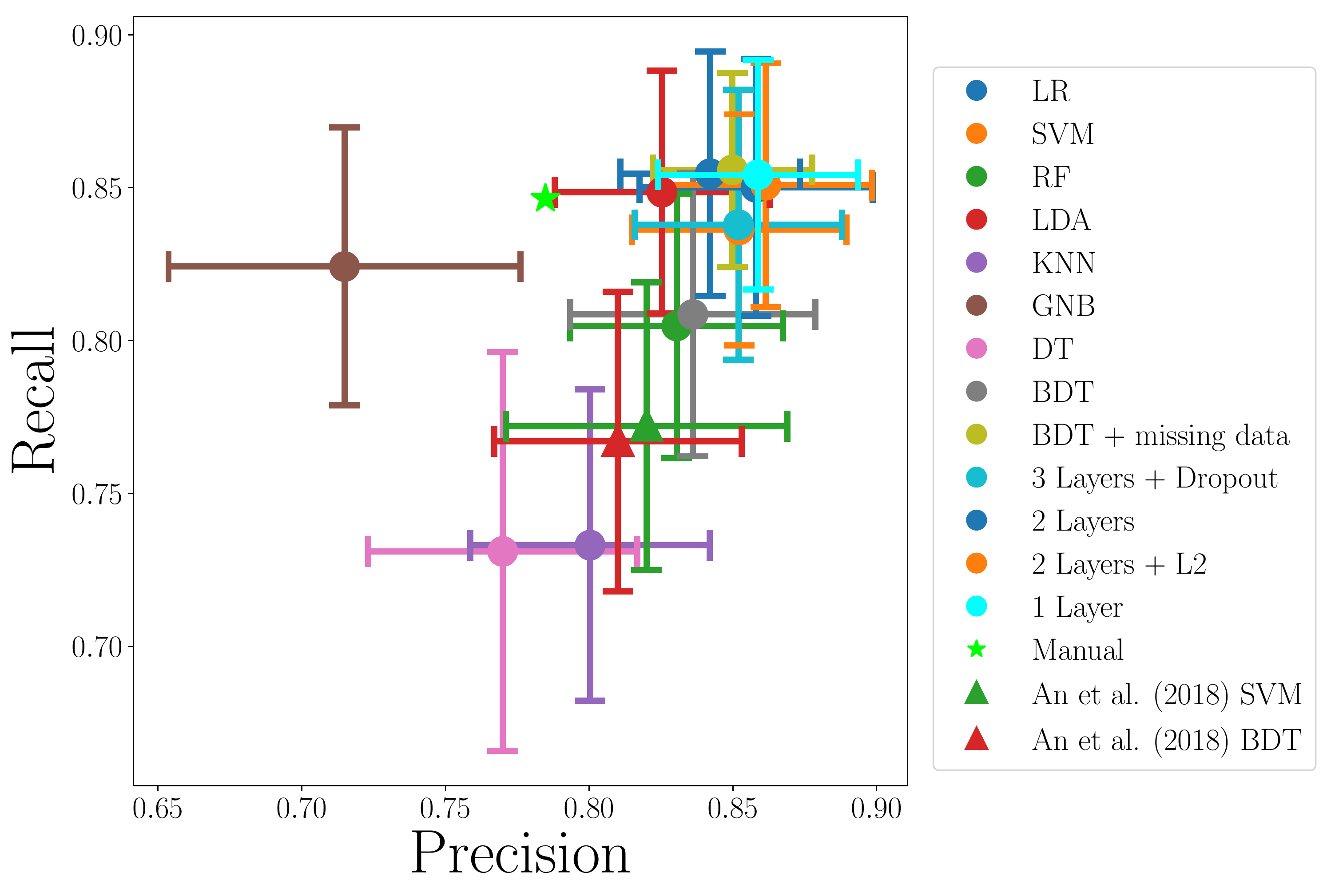}
  \caption{Precision and recall scores of different classifiers from cross-validation testing. Also included are the cross-validation results from \citet{an2018} using an SVM and a BDT. Most of the classifiers are concentrated near the top-right corner, with small differences on classification performance.}
  \label{fig:PrecisionRecall}
\end{figure}

The $k$-fold cross-validation metrics from the ML algorithms (trained without imputation) are shown in Table~\ref{tab:Classification Results}, and a plot of the precision/recall scores of all tested classifiers are shown in Fig.~\ref{fig:PrecisionRecall}. The results show that among the tested models, LR and the NN work best on our given data, while some other models (such as GNB and KNN) give rather poor performance. We also show the precision and recall reported in \citet{an2018} using an SVM with five features. By utilizing a smaller search radius and all the available features within our training set, our SVM and NN models were able to noticeably improve the performance compared to \citet{an2018}.

We further tested the false-positive rate of our NNs using the blank field approach. Our tests showed a false-positive rate (calculated as the number of SMGs identified in the blank fields over the total number of blank field targets) of about $3$ per cent. For comparison, Table~\ref{tab:Classification Results} shows that the precision of the NN is about 85 per cent.

It is worth noting that our manual classification algorithm was very competitive, with its results coming within only a few per cent in precision and recall compared to the much more complex models. However, despite the fact that the handful of bands shown in Figure \ref{fig:manualcut} appear to be well separated in colour space, it is still difficult to determine the optimal color combinations to use in the case of numerous bands without intensive trials. In this aspect, the benefits of the ML algorithms become their ability to provide a systematic procedure to fit the data in every parameter dimension. None the less, for this particular problem, incorporating complex models like deep learning, although beneficial, do not dramatically improve existing `manual learning' techniques in classification accuracy. An advantage of using a NN here might be that it is quick to implement (now that the software is readily available), only requiring some tuning of the number of layers and regularization. On the other hand, a disadvantage might be that, as a `black box', it would be difficult for the user to learn what features are providing the discriminatory power (in this case, redder colours), and this might be useful information for interpreting the data and hence planning for future observations.


In addition to regular cross-validation methods, we also tested our classifier models by training on all low S/N data and a portion of the high S/N data, while predicting on the remaining high S/N data. We found that this strategy had almost no effect on our performance metrics. Even when classifying high S/N sources, the classifiers' performances matched those of the original results within one standard deviation. This suggests that our training set is robust even when given relatively noisy photometry.



\begin{table}
	\centering
	\caption{Machine-learning performance calculated via $k$-fold cross-validation. We show the resulting mean and standard deviation for 200 random cuts of the data in this validation process. Note that $k$-fold cross-validation cannot be applied to the manual classification scheme, so we only report the metrics evaluated from a single run.}
	\label{tab:Classification Results}
    \begin{tabularx}{\columnwidth}{cYYYY}
    \toprule
    Algorithm & Accuracy & Precision & Recall & $F_1$ Score \\
    {} & [\%] & [\%] & [\%] & [\%]\\
    \midrule
    \multicolumn{5}{c}{ML algorithms} \\
    \midrule
    LR  & $92.0 \pm 1.3$ & $84.2 \pm 3.1$ & $85.5 \pm 4.0$ & $84.8 \pm 2.5$ \\
	SVM & $91.9 \pm 1.3$ & $85.2 \pm 3.7$ & $83.6 \pm 3.8$ & $84.3 \pm 2.6$ \\
	RF  & $90.6 \pm 1.5$ & $83.0 \pm 3.7$ & $80.5 \pm 4.3$ & $81.6 \pm 3.0$ \\
	LDA & $91.3 \pm 1.5$ & $82.5 \pm 3.7$ & $84.9 \pm 4.0$ & $83.6 \pm 2.9$ \\
    KNN & $88.3 \pm 1.7$ & $80.0 \pm 4.2$ & $73.3 \pm 5.1$ & $76.4 \pm 3.7$ \\
    GNB & $86.6 \pm 2.6$ & $71.5 \pm 6.1$ & $82.4 \pm 4.5$ & $76.3 \pm 3.8$ \\
    DT  & $87.2 \pm 2.0$ & $77.0 \pm 4.7$ & $73.1 \pm 6.5$ & $74.8 \pm 4.2$ \\
    BDT & $90.8 \pm 1.6$ & $83.6 \pm 4.3$ & $80.9 \pm 4.6$ & $82.1 \pm 3.3$ \\
    \makecell{BDT (with \\ missing data)} & $89.8 \pm 1.5$ & $85.0 \pm 2.8$ & $85.6 \pm 3.2$ & $85.2 \pm 2.2$ \\
    Manual & 90.2 & 78.5 & 84.6 & 81.4 \\
    \midrule
    \multicolumn{5}{c}{NNs} \\
    \midrule
    \makecell{3 Layers $+$ \\ Dropout}& $91.9 \pm 1.4$ & $85.2 \pm 3.6$ & $83.8 \pm 4.4$ & $84.4 \pm 2.8$ \\
    2 Layers & $92.4 \pm 1.4$ & $85.8 \pm 4.1$ & $85.0 \pm 4.2$ & $85.3 \pm 2.7$ \\
    \makecell{2 Layers $+$ \\ L2} & $92.5 \pm 1.4$ & $86.1 \pm 3.7$ & $85.1 \pm 3.7$ & $85.5 \pm 2.7$ \\
    1 Layer & $92.5 \pm 1.4$ & $85.9 \pm 3.5$ & $85.4 \pm 3.7$ & $85.6 \pm 2.7$ \\
    \bottomrule \noalign {\smallskip}
    \end{tabularx}
\end{table}

\subsection{Performance of imputation}
\label{sec:Impute_results}

Our preliminary tests with different imputation methods showed that the majority of the possible imputation algorithms were not useful. Simple methods, such as replacing the missing data with the mean or median, even performed worse in some cases. However, we did find that the Soft Impute \citep{Mazumder2010} model was promising. Hence we tested this algorithm more intensively by performing the same $k$-fold cross-validation as for the missing data on the most successful models discussed in Section~\ref{sec:CrossVal_results}. The resulting evaluated metrics are presented in Table~\ref{tab:SoftImpute Results}. 

Even with this specific imputation method, we saw no significant improvement compared to removing examples with missing data. However, the imputation tests at least showed us that a classifier trained with imputed data will not perform worse than a classifier trained only with real data. It is likely that we see almost no change in performance with imputation because our training set is large enough that, even after removing samples with missing data, our algorithms are able to learn everything they need to know to classify SMGs effectively. With smaller training sets, it is possible that imputation would be more important. Due to the negligible increase in performance of a model trained with the imputed data set, we chose to utilize only full-featured data for training when classifying new sources in the GOODS-N catalogue (see Section \ref{sec: GOODSN_SMGs}).

\begin{table}
	\centering
	\caption{Machine-learning performance using an imputed training set. The results of this test are unexceptional, since the metrics are mostly the same as the metrics obtained from training on data with no imputation.}
	\label{tab:SoftImpute Results}
    \begin{tabularx}{\columnwidth}{cYYYY}
    \toprule 
    Algorithm & Accuracy & Precision & Recall & $F_1$ Score \\
    {} & [\%] & [\%] & [\%] & [\%]\\
    \midrule
    \multicolumn{5}{c}{ML algorithms} \\
    \midrule
    LR & $91.7 \pm 1.5$ & $82.7 \pm 4.0$ & $86.5 \pm 3.9$ & $84.5 \pm 2.7$ \\
    BDT & $91.2\pm 1.5$ & $83.0 \pm 4.1$ & $83.5 \pm 4.0$ & $83.1 \pm 2.9$ \\
    Manual & 88.0 & 80.2 & 86.7 & 83.3 \\
    \midrule
    \multicolumn{5}{c}{NNs} \\
    \midrule
    \makecell{3 Layers $+$ \\ Dropout} & $91.9 \pm 1.4$ & $83.8 \pm 3.9$ & $85.2 \pm 4.2$ & $84.4 \pm 2.8$ \\
    2 Layers & $92.2 \pm 1.5$ & $83.8 \pm 4.3$ & $86.7 \pm 3.7$ & $85.2 \pm 2.9$ \\
    \makecell{2 Layers $+$ \\ L2}& $91.9 \pm 1.4$ & $83.7 \pm 3.9$ & $85.7 \pm 4.0$ & $84.6 \pm 2.6$ \\
    1 Layer & $92.3 \pm 1.4$ & $84.1 \pm 3.9$ & $87.0 \pm 3.3$ & $85.4 \pm 2.6$ \\
    \bottomrule \noalign {\smallskip}
    \end{tabularx}
\end{table}

\subsection{Receiver operating characteristic curves}
\label{sec:ROC}

\begin{figure}
  \includegraphics[width=\columnwidth]{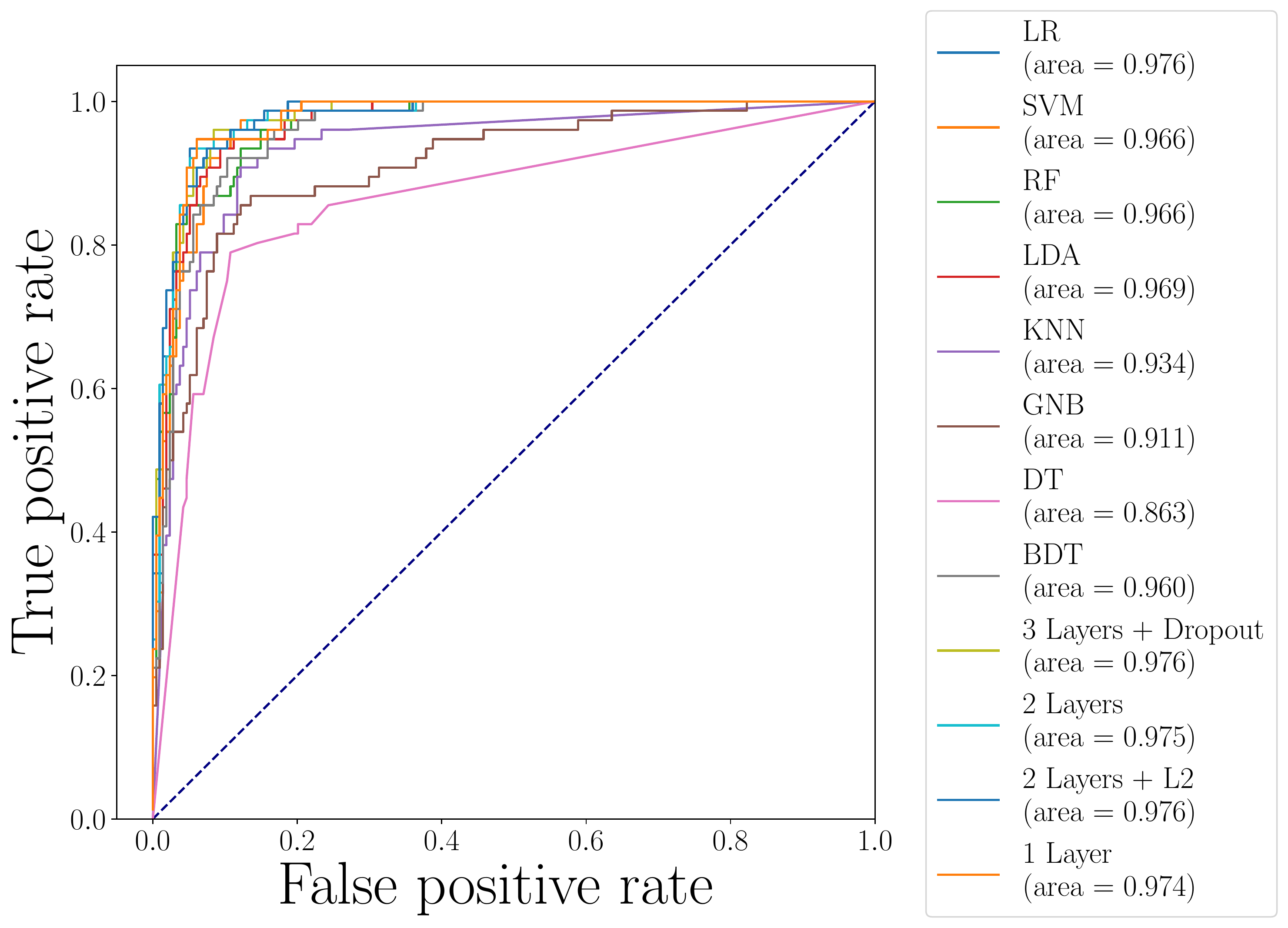}
  \caption{ROC curves for a series of classifiers using the non-imputed data.}
  \label{fig:ROC}
\end{figure}

Another metric that can be used to evaluate classifiers is the receiver operating characteristic (ROC) curve. This is a plot of the true positive rate against the false positive rate at different thresholds of the classifier (i.e. different dividing points for classifying a source as an SMG or a non-SMG), effectively evaluating the classifier at all possible thresholds. The effectiveness of the ROC curves can be evaluated by calculating the area-under-curve (AUC) score for each of these functions. As a general rule, the closer a curve gets to the upper left corner of the plot (and hence the larger the AUC), the better the classifier. We plot such an ROC curve for all of our tested classifiers in Fig.~\ref{fig:ROC}, and calculate the areas (which we include in the figure legend).

\subsection{Model selection}

We selected the best classifier based on the classification results in Tables~\ref{tab:Classification Results} and \ref{tab:SoftImpute Results}, as well as Figs.~\ref{fig:PrecisionRecall} and \ref{fig:ROC}. The four NNs performed best in three of the four measured metrics among the classifiers we tested, and they had the best AUC score in the ROC plot. Other classifiers, such as LR and the SVM, also performed relatively well, but fell short of the NNs. 
Overall, the most successful NN is the 1-layer NN. The 1-layer network is only slightly superior to other NN models we tested, even falling slightly behind on some metrics such as the ROC score (although the difference of 0.002 in the ROC score is quite small). Ultimately, we selected our preferred network based on the $F_1$ score because we believe it to be the best judge for overall model performance in this particular application. 

\vspace{0.5cm}
\section{Application to the GOODS-N field}
\label{sec:NewPred}

\subsection{A catalogue of ML-identified SMGs}
\label{sec: GOODSN_SMGs}

Using the most successful 1-layer NN model, we classified the sources in the S2CLS found within the GOODS-N field described in Section~\ref{sec:PredGOODSN}. The matched catalogue contains 290 full-featured multiwavelength sources, corresponding to 67 SCUBA-2 beam areas in the GOODS-N field. Identifying SMGs without the use of high-resolution submm data is particularly important in this field since, being in the northern hemisphere, ALMA is unable to observe it.

Although the 1-layer NN gave good results, the nature of the NN with random weight initialization means that trained models still have some intrinsic variance. When training and predicting on this field with a single NN classifier, we found that each successive run yielded a different number of positively classified SMGs, betweeen 40 and 50.
A catalogue classified from a single NN is thus problematic due to this high variance, and we would like to have a definitive result unchanged by each run. To overcome this problem, we utilised a neural-network ensemble-averaging method \citep{Hansen1990}, also known as bagging, for our predictions. By training a large number of NNs (each with different random weight initializations) in parallel and averaging the returned results, we can find the typical spread in the classification. Furthermore, by taking into account the standard deviation of the returned prediction scores from each individual NN, we would be able to quantify uncertainties for our predictions \citep{Lakshminarayanan2016}.



We applied this ensemble bagging algorithm to the GOODS-N catalogue by training 100 NNs in parallel and taking the average and standard deviation of the resulting prediction scores. 
We employed a cutoff at 0.5 for the output averaged scores, classifying all sources with a score above this point as SMGs and all of those below as non-SMGs (Appendix \ref{app:DNN} describes in detail how the prediction score is generated).
    
We found a total of 45 positive matches within the 290 full-featured sources. These 45 multiwavelength ML-identified SMG candidates lie in 36 SCUBA-2 sources (out of a possible 67), with eight SCUBA-2 sources having more than one multiwavelength candidates. We see that our prediction catalogue contains eight SCUBA-2 sources with multiple IDs (seven of which are matched to two SMG candidates, while one SCUBA-2 source is matched to three). For comparison, the SMA follow-up of \citet{cowie2017} found three pairs of SMGs in the GOODS-N field, comparable to our result given the large Poisson error bars, while the ALMA follow-up (used as our training set) contained many more pairs of SMGs due to its much higher sensitivity and angular resolution.  A table of all the classified source IDs within GOODS-N can be found in Appendix~\ref{app:GDN_SMG}. To aid further studies and independent confirmation, we give the most relevant names and identification numbers, along with NN prediction scores for each of the sources. It should be emphasized that the NN prediction scores do not represent actual probabilities, but instead a potentially non-linear certainty for a given classification. No SMG identifications were made for 22 of the SCUBA-2 sources. This could be because these SMGs are simply too faint to be detected in the multiwavelength catalogue (as for HDF850.1, see below).
    

We ran a blank field test (slightly different from the blank field test in Section \ref{sec:Validation}) focused on all sources within 7\,arcsec `beams' in the GOODS-N field by taking random RA and Dec positions within the field and running our trained NN on the multiwavelength sources contained within the simulated beams. From such a test, we found that only 37 out of 1000 fake `beams' contained any false positive IDs, giving a false positive of 3.7 per cent. Comparing this to our results on real SCUBA-2 sources, where we found that 36/67 (or about 55\,per cent) of the SCUBA-2 sources in the GOODS-N field matched with counterparts in the multiwavelength catalogue, we see that the false positive rate of our classifier is quite low.

It is also worth noting once more that as the two fields (UDS and GOODS-N) used for training and predicting are not identical (in terms of depth, source extraction, etc.), the accuracy of the NN in this application may be different than what we what we estimated from our extensive testing. For instance, as discussed in Section \ref{sec:PredGOODSN}, we found that the depths of the two catalogues differ by a factor of about 2. We tested the effects of this difference by removing faint sources from each catalogue such that their minimal 3$\sigma$ depths matched one another. This reduced our training set from 301 to 145 SMG sources and 856 to 537 non-SMGs. Likewise, the size of the prediction set was reduced from 290 fully-featured objects to 173. Using the same training and predicting methods described above, we found that the set of neural networks trained on this reduced data yielded similar results compared to using the full data; in particular, we found that the mean absolute difference in the output prediction scores resulting from these two training sets was about 3\,per cent. Since this difference is quite small, we are confident that we can use the full depth available in both fields in order to utilize a larger training set and classify more GOODS-N sources.

\subsection{Comparison with known GOODS-N SMGs}
\label{sec:ComparisonGOODSN}

To test our results from an astronomical perspective, we checked a couple of well-known SMGs in the GOODS-N field, specifically the galaxies known as GN20 \citep{pope2005} and HDF850.1 \citep{hughes1998}. With the former, we were able to locate six multi-wavelength sources within the SCUBA-2 beam (four of them with no missing features), and we knew a prior which one of these sources corresponded to the actual submm source. Predicting with the ensemble algorithm, we were able to classify the correct multiwavelength source with a prediction score of 0.7, while the other galaxies in the beam were all correctly classified as non-SMGs. In particular, we found one non-SMG source within the beam area that had similar features compared to the real GN20 source. Although this source seemed similar by eye, the NN accurately classified it as a non-SMG, with a prediction score of 0.15, well within our 0.5 cutoff.
    
For the other well-studied SMG, HDF850.1, multiwavelength identification has been a long battle \cite[see e.g.][]{walter2012}, with the current consensus being that the source is extremely faint at optical and NIR wavelengths; indeed, our current GOODS-N catalogue does not contain HDF850.1. Nevertheless, we were able to confirm that no false SMGs were found within the SCUBA-2 beam surrounding HDF850.1.

The GOODS-N field has also seen submm interferometric follow-up with with the SMA from \citet{cowie2017}. As an independent test, we compared our ML identifications of 36 SCUBA-2 sources with the 33 SMA detections of SCUBA-2 sources found in their work; out of these 33 detections, 15 matched with our identifications, four were sources that our NN deemed not to be SMGs, and three were matched to sources in our multiwavelength catalogue with missing photometry that we did not run our NN on. The remaining 11 were not matched to any source from our catalogue. This is equivalent to a matching success rate of 68 per cent and a completeness of 45 per cent.
    
From this comparison, it is clear that although our classifiers do match with known sources from other catalogues, there is a disadvantage with only predicting on sources with full multiwavelength data. 
In particular, some bright SMGs could only be bright enough to be detected in the NIR, and lack optical photometry detection altogether. In our GOODS-N-matched catalogue, only 290 of the 399 possible galaxies are fully featured, meaning we were not able to make a classification for over a quarter of the matched data set.
In order to increase the number of sources classified, two approaches could be used. The first is to find additional features more relevant for these sources, which we discuss in Section \ref{sec:radio24}. The second is to fill in the missing features with, for example, aperture photometry, even if bands may be below 3$\sigma$ detections (discussed in Section \ref{sec:missingfeatures}).


\subsection{The Super-deblend catalogue of the GOODS-N field}

The GOODS-N field has seen other attempts at untangling the FIR to mm wave images \citep[e.g.][]{borys2004, chapin2009, pope2006, gruppioni2013}. Recently, \citet{liu2018} developed a technique called `super-deblending', which uses {\it Spitzer\/} 24-$\mu$m and Karl G.~Jansky Very Large Array (VLA) 20-cm imaging (both of which have good angular resolution) to create a base catalogue of galaxies as candidates for the submm emission seen by {\it Herschel\/}-SPIRE at 250, 350 and 500\,$\mu$m. The technique then looks at a series of images, starting from 24\,$\mu$m and moving to progressively longer wavelengths where the resolution deteriorates, sequentially removing candidate galaxies deemed hopelessly faint in the submm based on the best-fitting model SED at a given stage. This work is similar to our own in that it tries to identify SMGs from catalogues of galaxies where the resolution is far superior. However, key differences include the fact that the super-deblend algorithm requires an assumed SED shape (whereas our NN `learns' what a good SED shape is), and that its primary task is to estimate the flux density of blended galaxies (as opposed to identifying their counterparts).


As a check of our GOODS-N catalogue, we compared our results to those of \citet{liu2018}. In order to do this, we took all sources in the super-deblend data set that were predicted to be brighter than 0.89\,mJy in the 850\,$\mu$m band. This threshold was chosen because it corresponds to the noise limit of ALMA-detected sources in our training set \citep{Stach2018}. We then matched these sources to the SCUBA-2 beam areas in the GOODS-N field. We found that in the super-deblend data set, there were 63 sources meeting our brightness criteria that matched within a 7\,arcsec radius of our SCUBA-2 sources. Of these sources, 38 also matched with our full-featured multiwavelength catalogue to within 0.6\,arcsec; these 38 sources should therefore presumably be classified as SMG candidates by our best ML algorithm. Upon comparing these 38 sources to our ML results, we found that our algorithm identified 33 as SMG candidates. In addition, our GOODS-N catalogue contains 12 candidates that were not found with super-deblending, while the super-deblending catalogue contains five candidates that were not found here. 

There could be a variety of reasons for the discrepancies between the two catalogues. For example, one of the five sources we identified as a non-SMG was given a 0.4995 prediction score of being an SMG by the NN, barely missing the cutoff. Three other sources were matched with very faint SCUBA-2 sources at the bottom end of the S/N limit. Similar issues might occur in the super-deblend data set, and in practice neither of the two catalogues will give definitive results for SMGs, since we are operating at only modest S/N values. Nevertheless, there is reassuringly good overall agreement between our two approaches.

\section{Future improvements}
\label{sec:future}

\subsection{Combining deblending and machine learning}
\label{sec:DeblendML}

In the future, it might be possible to combine deblending and ML approaches, using the power of a NN to identify the useful candidates and the power of deblending to obtain the photometric properties of the galaxies in the submm. Indeed, the super-deblend method described above is only one implementation of a more general idea; for example, a similar method called `SEDeblend' \citep{mackenzie2014,mackenzie2016} uses a high-resolution catalogue to simultaneously fit a number of long-wavelength images while also fitting the SEDs of the blended sources. As previously mentioned, a fundamental difference of these kinds of methods are that they are {\it fitting\/} the images, rather than simply deciding whether each catalogue source is a good SMG candidate.

Combining deblending and ML would require using an iterative scheme with several training sets in different wavebands, so that at each step the decision between useful versus hopelessly faint could be made more accurately. However, this still leaves open the question of how to connect the model SED (which is needed to obtain a predicted flux density) to the SED `learned' by a NN. Further work is required to investigate how to extract results from a NN when given an SED to fit. Another potential method for combining the techniques might involve using deblending to fill in missing wavelengths in the multiwavelength catalogues in order to train and classify bigger catalogues with ML algorithms.

\subsection{Using radio and \texorpdfstring{24$\,\mu$m catalogues}{}}
\label{sec:radio24}

There is of course no reason to restrict the training set to short wavelengths -- as long as the resolution is sufficient to keep individual galaxies from becoming too blended, longer wavelengths (more similar to the submm regime) should provide useful features for SMG identification. \citet{an2018} in particular used the radio IDs of SMGs in the UDS field to find multiwavelength counterparts for sources where their ML algorithm was not able to find one on its own; this was found to increase the recall by about 5\,per cent.

As an exploratory test of this idea, we compared our catalogue of GOODS-N identifications to surveys done in the radio using the VLA \citep{morrison2010} and at 24\,$\mu$m using {\it Spizter},\footnote{\url{https://irsa.ipac.caltech.edu/data/SPITZER/GOODS/docs/goods_dr1plus.html}}$^,$\footnote{\url{https://irsa.ipac.caltech.edu/data/SPITZER/GOODS/docs/goods_dr1plus_mipslist.html}} with the aim of quantifying how many GOODS-N SCUBA-2 sources left unaccounted for by our NN might benefit from the incorporation of these additional data. To find the number of SCUBA-2 sources with potential radio/24\,$\mu$m counterparts, we used the matching criteria outlined in \citet{downes1986}, which takes into account a finite survey depth and search radius. Here a $p$-value is computed for all radio/24\,$\mu$m sources within a search radius of each SCUBA-2 source (taken to be 7\,arcsec to match the FWHM of the SCUBA-2 data). First, we found that out of the 36 SCUBA-2 sources already given a multiwavelength ID by our NN, 23 also had a radio counterpart and, independently, another 23 had a 24\,$\mu$m counterpart. Second, out of the remaining 31 SCUBA-2 sources lacking a ML-identified counterpart, 13 had a radio or 24\,$\mu$m association. For completeness, in Table \ref{tab:predGDNs} we also show which SCUBA-2 sources have radio IDs and 24\,$\mu$m IDs.


The important question then is whether adding radio and 24$\mu$m data as features would be beneficial to our classification algorithm. From our results, it is clear that if we consider the matched radio and 24\,$\mu$m sources as positive identifications, our identification rate would increase from 36/67 to 49/67. This means that by adding this data we could potentially increase the total fraction of identified SCUBA-2 sources to 73 per cent.  However, due to the lack of mutual multiwavelength and radio/24\,$\mu$m data for a lot of sources, we would need to implement one of the imputation techniques described above to include sources with missing features properly, or come up with a new way to construct the list of feature.

\subsection{Missing Features and uncertainties}
\label{sec:missingfeatures}

Currently, our classification algorithm does not take into account missing features or flux density uncertainties. This limits us to classifying only a subset of all total detected sources in a given field, while also treating sources with different S/N in the same manner. Although our classifier is still shown to be effective and matches other results, we would like in future studies to incorporate both missing featured sources and photometric uncertainties, in order to produce a more robust and accurate catalogue.

Missing features may be due to a variety of reasons. The coverage of surveys, sources too faint to pass a S/N threshold, and nearby bright sources could all lead to missing photometric magnitudes in catalogues. In particular, some bright SMGs are only bright in the submm-to-infrared range, and intrinsically lack the brightness in optical wavelengths to be detected by photometric surveys. These cases are particularly interesting because they are clear hints that an SMG is present. In the future it may be more helpful to train a separate ML algorithm using only photometry from the redder wavebands, along with the possible inclusion of radio and 24\,$\mu$m catalogues, as described in Section \ref{sec:radio24}.

To solve the problem of sources with missing survey coverage, we have already seen the potential of imputation for helping with missing features. Section \ref{sec:Impute_results} showed that an imputed training data set would perform no worse than a normal training data set when tested on real data. But it is worth noting that imputing a prediction set may lead to systematic biases that are not corrected for by the classifier, and that these biases are likely to give misleading classification results. Therefore, systematic testing will need to be done to observe the effectiveness of creating new artificial data in the prediction set for classification.

A second method for filling in missing features relies on having aperture photometry at the positions of known red (i.e.~bright at longer wavelength) sources, even if the flux is consistent with zero or negative. In this way, galaxies that are essentially invisible at optical wavelengths would not have to be removed from the ML training and predicting sets, and ML algorithms could `learn' that these sources will probably be bright in the submm.

Uncertainties pose another potential problem for classification. The photometric magnitudes used as features in our algorithm have systematic and statistical uncertainties related to their original measurements in surveys. At the moment, these uncertainties are not taken into account by the algorithm, and we are only classifying each source based on the measured value of the magnitudes in the catalogues. However, it would be possible to factor in the S/N of our multiwavelength sources when classifying in the future. One method of incorporating uncertainties is to sample our features from their error distributions, rather than taking the measured values. We could use a similiar ensemble bagging system to what was described in Section \ref{sec: GOODSN_SMGs}, either training or predicting each individual NN with a slightly different data set sampled from the error distributions, and averaging the final results \citep{Hansen1990}. By sampling in an ensemble of classifiers, we would be able to classify each source multiple times based on their error distribution, and average the results to obtain a collective classification for each source. However, it is important to note that such a large scale ensemble would require significant computational resources.

Since both our training catalogue and any potential prediction catalogue will contain uncertainties, further testing is also required to understand the effects of such an ensemble method. We would need to sample both a training set before training a model, and sample prediction sets before predicting on a new catalogue.



\section{Conclusions}
\label{sec:Conclusions}

Based on our rigorous testing of various ML algorithms trained using data from the UDS field, a subset of the S2CLS, it is clear that an ML approach is a useful way to classify potential counterparts to SMGs. Our testing of different algorithms showed us that a NN model performs best compared to other ML algorithms in this application. However, as we have shown with our manual colour-cut comparison, ML techniques do not always provide dramatic improvements over more traditional methods as one might hope. Nevertheless, the the additional 5\,per cent in accuracy is worth the extra effort, since it comes without the need to gather any extra data.



We then applied our ML algorithm to the GOODS-N field, another subset of the S2CLS that contains similar submm and multiwavelength data but lacks a thorough interferometric follow-up programme (with e.g.~ALMA). We identified counterparts for 36 out of the 67 submm sources, and our classifications roughly matched with a number of other attempts to pinpoint the locations of SMGs within the field, including a partial survey of the field by the SMA.

To conclude, it is worth pointing out that although ML can out-perform traditional methods, one has to be careful to perform a fair comparison. There are certainly disadvantages to ML techniques, especially the most advanced methods.  First, it can be dramatically more difficult to set up the analysis pipeline, which then requires more computational resources. Second, it is usually hard to determine uncertainties, or to interpret the uncertainties that are produced by the data-science codes.  And finally, it can be challenging to determine what underlying features are being used to make the decisions within the ML process and thus learn something useful about the sources being studied.



\section*{acknowledgements}

We would like to acknowledge the Natural Sciences and Engineering Research Council of Canada for making this research possible. SMS acknowledges the support of STFC studentship (ST/N50404X/1). IRS acknowledges support from the ERC Advanced Investigator programme DUSTYGAL 321334 and STFC grant (ST/P000541/1). The James Clerk Maxwell Telescope is operated by the East Asian Observatory on behalf of The National Astronomical Observatory of Japan; Academia Sinica Institute of Astronomy and Astrophysics; the Korea Astronomy and Space Science Institute; the Operation, Maintenance and Upgrading Fund for Astronomical Telescopes and Facility Instruments, budgeted from the Ministry of Finance (MOF) of China and administrated by the Chinese Academy of Sciences (CAS), as well as the National Key R\&D Program of China (No. 2017YFA0402700). Additional funding support is provided by the Science and Technology Facilities Council of the United Kingdom and participating universities in the United Kingdom and Canada. The SCUBA-2 data presented in this paper was taken as part of Program ID MJLSC02. This paper makes use of the following ALMA data: ADS/JAO.ALMA\#2012.1.00090.S, 2015.1.01528.S, and 2016.1.00434.S. ALMA is a partnership of ESO (representing its member states), NSF (USA) and NINS (Japan), together with NRC (Canada), MOST and ASIAA (Taiwan), and KASI (Republic of Korea), in cooperation with the Republic of Chile. The Joint ALMA Observatory is operated by ESO, AUI/NRAO and NAOJ. This work is based in part on observations made with the Spitzer Space Telescope, which is operated by the Jet Propulsion Laboratory, California Institute of Technology under a contract with NASA. Based on observations obtained at the Canada-France-Hawaii Telescope (CFHT) which is operated by the National Research Council of Canada, the Institut National des Sciences de l'Univers of the Centre National de la Recherche Scientifique of France, and the University of Hawaii. Based in part on data collected at Subaru Telescope, which is operated by the National Astronomical Observatory of Japan. The UKIDSS project is defined in \citet{lawrence2007}. Further details on the UDS can be found in Almaini et al.~(in prep). UKIDSS uses the UKIRT Wide Field Camera (WFCAM; \citealt{casali2007}). The photometric system is described in \citet{hewett2006}, and the calibration is described in \citet{hodgkin2009}. The pipeline processing and science archive are described in Irwin et al.~(in prep) and \citet{hambly2008}. The National Radio Astronomy Observatory is a facility of the National Science Foundation operated under cooperative agreement by Associated Universities, Inc.
\vspace{1cm}



\bibliographystyle{mnras}
\bibliography{MLCite}



\appendix

\section{Machine Learning Models}
\label{app:ML_models}

Here we give qualitative descriptions of the ML approaches used in this paper, as well as further references. For a more in-depth description of the Boosted Decision Tree, as well as the Python implementation, see the package \texttt{xgboost}.\footnote{\url{https://xgboost.readthedocs.io}} For the rest of the algorithms, see the package \texttt{scikit-learn}.\footnote{\url{ https://scikit-learn.org/stable/user_guide.html}} and the accompanying documentation.

\begin{itemize}
    \item {\bf Logistic regression} \citep[LR;][]{peng2002}: also known as `logit regression', is a linear classification model that linearly combines the data features into a single number using weights optimized during training. The algorithm then applies a nonlinear sigmoid function as a filter (see Eq.~\ref{eq:Sigmoid}), and this results in a number between zero and one. In this way, a logistic regression algorithm can be considered as a neural network with no hidden layers. 
        
    \item {\bf Support vector machine} \citep[SVM;][]{Cortes1995}: creates hyperplanes in $n$-dimensional feature space in order to separate the possible classifications; these hyperplanes are tuned during the training process in order to place as many training example features as possible into each classification section. Predictions are then made by locating new data within these sections and assigning the corresponding label.
    
    \item {\bf $k$-nearest neighbors} \citep[KNN;][]{Cover1967}: plots the training data in an $n$-dimensional parameter space, and simply classifies new data by finding the $k$ nearest neighbours based on their Euclidean distances and looking for the most frequent label among these $k$ neighbours. 
        
    \item {\bf Gaussian naive-Bayes} \citep[GNB;][]{Friedman1997}: assumes that each feature in a given training example is independent, and furthermore that within each classification features are drawn from a Gaussian distribution. Priors are set for each possible classification as the relative frequency of each class within the training set, and the means and variances of each Gaussian distribution are tuned during the training process. The classifier then uses Bayes' theorem to calculate the probability that the input samples belong to each label class, and takes the maximum probability as the output. 
    
     \item {\bf Linear discriminant analysis} \citep[LDA;][]{Ye2007}: assumes that all features come from multivariate Gaussian distributions with equal covariance matrices, regardless of their classification. Priors are set in the same way (as the relative frequency of each classification), means are calculated for the features within each classification, and the covariance matrix is tuned during training. The classifier then uses Bayes' theorem in the same way as the Gaussian naive-Bayes algorithm to assign classifications to new data.
    
    \item {\bf Decision trees} \citep[DT;][]{Quinlan1986}: use a tree-like structure to separate data using a number of different decisions thresholds that are tuned during the training process. For each decision the algorithm looks at one or more feature and checks to see where the feature falls among the thresholds, then based on the answer follows two or more possible branches; the number of decisions, the number of features to consider in each decision, and the number of threshold boundaries/branches are chosen beforehand, while the features used in each decision and the decision thresholds are optimized during the training. New data are then passed into the decision tree and filtered down through the branches based on the decision boundaries, and eventually into decision nodes that provide a classification.
    
    \item {\bf Random forests} \citep[RF;][]{Ho1995}: are similar to decision trees, but use an ensemble of uncorrelated decision trees to classify the data. Each decision tree in a random forest is trained on a subset of the whole training set, and the final classification looks at the output from each decision tree and identifies the most predicted result. The free parameters that are tuned during the training process are the same as with the decision tree algorithm, but here there are multiple sets of these parameters, based on the number of trees chosen to include in the random forest.
        
    \item {\bf Boosted decision trees} \citep[BDT;][]{friedman2001}:
    is a variant of the random forest algorithm. It uses an algorithm called `boosting' in which individual decision trees are correlated with one another and fit consecutively; thus there are additional parameters to tune during training, which are the correlation coefficients between the trees in the random forest. This model is implemented with the \texttt{xgboost} package.
    
\end{itemize}

\section{Deep Neural Networks}
\label{app:DNN}

\begin{table*}
	\centering
	\caption{Structures of our four tested deep learning neural networks. Networks with regularization can have more nodes and parameters to train, since the regularization step will help with overfitting issues. The details in brackets give the type of layer, e.g. the number of layers, activation method and specific parameters, (as explained in the text).}
	\label{tab:NNets}
    \begin{tabular*}{\textwidth}{c @{\extracolsep{\fill}} cccc}
    \toprule \noalign {\smallskip}
    Model & \makecell{3 layers $+$ \\ Dropout} & 2 layers & \makecell{2 layers $+$ \\ L2} & 1 layer\\
    \midrule
    Layer 1 & \makecell{Fully connected \\(64, ReLU)} & \makecell{Fully connected \\(32, ReLU)} & \makecell{Fully connected with \\ L2 (64, ReLU, L2 $\lambda = 0.001$)} & \makecell{Fully connected \\(32, ReLU)} \\
    Layer 2 & Dropout (25\%) & \makecell{Fully connected \\(4, ReLU)} & \makecell{Fully connected with \\ L2 (8, ReLU, L2 $\lambda = 0.001$)}& \dots \\
    Layer 3 & \makecell{Fully connected \\(64, ReLU)} & \dots & \dots & \dots \\
    Layer 4 & Dropout (25\%) & \dots & \dots & \dots \\
    Layer 5 & \makecell{Fully connected \\(8, ReLU)} & \dots & \dots & \dots \\
    Output Layer & \makecell{Fully connected \\ Output (1, Sigmoid)} & \makecell{Fully connected \\ Output (1, Sigmoid)} & \makecell{Fully connected \\ Output (1, Sigmoid)} & \makecell{Fully connected \\ Output (1, Sigmoid)}\\
    \midrule
    \makecell{No.\ of trainable\\ parameters} & 5,649 & 617 & 1,489 & 513 \\

    \bottomrule \noalign {\smallskip}
    \end{tabular*}
\end{table*}

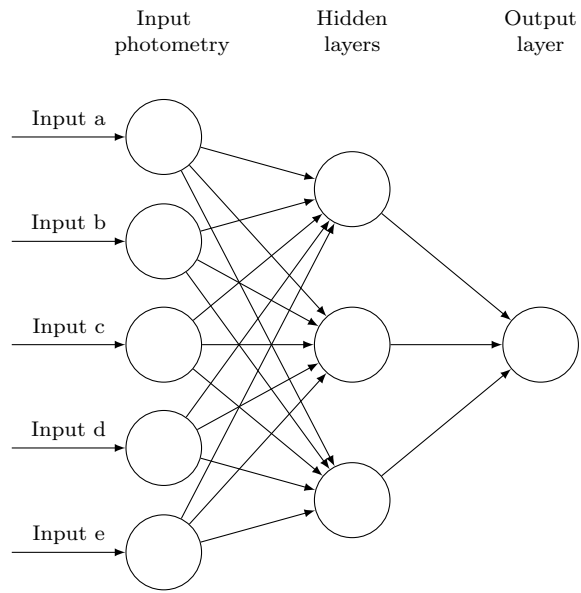
\begin{figure}
\centering
\begin{tikzpicture}[
x=1cm, 
y=1.5cm, 
>=stealth,
plain/.style={
  draw=none,
  fill=none,
  },
net/.style={
  matrix of nodes,
  nodes={
    draw,
    circle,
    inner sep=10pt
    },
  nodes in empty cells,
  column sep=0.1cm,
  row sep=-9pt
  },
>=latex
]
\matrix[net] (mat)
{
|[plain]| \parbox{1.3cm}{\centering Input\\photometry} & |[plain]| \parbox{1.3cm}{\centering Hidden\\layers} & |[plain]| \parbox{1.3cm}{\centering Output\\layer} \\
& |[plain]| \\
|[plain]| & \\
& |[plain]| \\
  |[plain]| & |[plain]| \\
& & \\
  |[plain]| & |[plain]| \\
& |[plain]| \\
  |[plain]| & \\
& |[plain]| \\    };
\draw[<-] (mat-2-1) -- node[above] {Input a} +(-2cm,0);
\draw[<-] (mat-4-1) -- node[above] {Input b} +(-2cm,0);
\draw[<-] (mat-6-1) -- node[above] {Input c} +(-2cm,0);
\draw[<-] (mat-8-1) -- node[above] {Input d} +(-2cm,0);
\draw[<-] (mat-10-1) -- node[above] {Input e} +(-2cm,0);

\foreach \ai in {2,4,...,10}
{\foreach \aii in {3,6,9}
  \draw[->] (mat-\ai-1) -- (mat-\aii-2);
}
\foreach \ai in {3,6,9}
  \draw[->] (mat-\ai-2) -- (mat-6-3);
\end{tikzpicture}
\label{fig:nnetdiagram}
\caption{A visual representation of a typical neural network architecture. Similar network architectures are used in this work to decide between SMGs and non-SMGs.}
\end{figure}

DNNs are a popular type of algorithm for tackling ML and data-science problems \citep{Schmidhuber2014}. A DNN is made up of interconnected layers of neurons, as shown in Figure~\ref{fig:nnetdiagram}. Each neuron $j$ performs a linear combination on an input array $\vec{x}$ from the layer before it by applying to $\vec{x}$ an array of trainable weights $\vec{w}^j$, as well as adding a bias $b^j$. The values of $\vec{w}^j$ and $b^j$ are specific to each neuron, and set during network training. The neural network then implements a non-linear `activation function' $f$ on the result, giving the final output of the neuron. The choice of activation functions is a changeable hyperparameter of the neural network. In our case, we used a rectified linear unit (ReLU) function, defined as
\begin{equation}
\label{eq:RELU} 
f(x)=\max(0,x),
\end{equation}
for the activation function in the hidden layers.

The output layer result is normalized instead with a sigmoid activation function, defined as
\begin{equation}
\label{eq:Sigmoid}
S(x) = \frac{1}{1+\exp(-x)}.
\end{equation}
The sigmoid function serves to normalize the output of the network between 0 and 1, and the output will essentially act as a `probability' of the classified source being an SMG (1) or non-SMG (0); note that the score does not symbolize a real probability for the source being a `SMG', but rather a quantified certainty value for the class given by the neural network, the higher the value, the more certain the neural network is that this source is of this class. We choose The complete operation of each neuron $y_j$ is shown in Equation~\ref{eq:NNet}, where $n$ is the number of weights connecting to the neuron:
\begin{equation}
\label{eq:NNet}
y_j = f(\,\vec{x}\cdot \vec{w}^j+b^j\,) = f\left(\, b^j+\sum_{i=1}^{n}\; x_i w^j_i\,\right).
\end{equation}


Typically, the first layer of a neural network is the input, and the last the output, with a number of hidden layers between. When training, the algorithm will take the input data and pass it through the neural network, and a loss function will quantify the difference between the end result and the expected result. In a binary classification problem, such as the classification of SMGs, where the expected output value is a float between 0 and 1 (with 0 being the negative class and 1 being the positive class), a very common loss function is binary cross-entropy. Binary cross-entropy loss is defined as 
\begin{equation}
\label{eq:BCent}
H(y, \hat{y}) = -y\log(\hat{y}) + (1-y)\log(1-\hat{y}),
\end{equation}
where $y$ is the true binary label (i.e.~either 0 or 1), and $\hat{y}$ is the predicted `probability' label. This loss function will be optimized by determining the best weights in the network. Using an optimization algorithm \citep[usually some variant of stochastic gradient descent][]{bottou2004}, the neural network will then change the weights based on the loss function and find the best possible set of weights to connect the input layer to the output. Neural networks essentially behave as complex non-linear functions that may be more sensitive to the relevant patterns than less complex algorithms. 

A common problem in neural-network training is `overfitting' \citep{Buduma2015}.  This occurs when a complex model becomes too sensitive to the input training data and is unable to generalize on new testing data. Rather than `learning' the data, an overfitted model will tend to `memorize' the data, resulting in near perfect training scores, but a much lower testing score on a separate validation set. To reduce overfitting, a number of different methods exist. The simplest approach is to reduce the complexity of the model. Neural network models rarely need to go beyond two or three layers, and larger neural networks will tend to overfit. An alternative to reducing the number of nodes in a model is to use regularization techniques to combat overfitting. 
One of the most commonly used regularization techniques, L2, works by adding the following term to the loss function:
\begin{equation}
   \frac{\lambda}{2n} \sum_{i} || w_{i} ||^2.
   \label{eq:L2}
\end{equation}
Here $w_{i}$ represents a single weight in the network and $n$ represents the number of samples, while $\lambda$ is a hyperparameter that can be optimised. A third way to reduce overfitting is by using  the dropout technique \citep{Srivastava2014}; this is a recent and popular regularization method that is implemented by randomly selecting a number of nodes in a layer and setting them to zero. This forces the neural network to adapt to missing certain information and prevents the dependence on any particular node. In our case, we attempted to integrate all these techniques in different neural network models in order to find the best approach for our data.

In a complex algorithm, such as a neural network, there are many parameters that need to be chosen. As discussed in Section~\ref{sec:Hyperparams}, we have fixed some of the parameters before optimising for the rest. However, we still decided to optimise some specific training parameters, such as epochs and batch sizes. An epoch is an iteration of passing the whole data set through the network for a feed-forward neural network such as our own (i.e.~a network that passes through data in only one direction without forming a cycle); in practice multiple epochs are needed in order to achieve a good result. A batch size is the number of samples to pass through the neural network before making an adjustment to the weight parameters. Through our optimization algorithm, we found that training each network for 150 epochs, with 64 samples per batch provided a good solution.

In addition to the epoch and batch size, we also tested different optimiser algorithms for the neural network. We tested two specific algorithms, Stochastic Gradient Descent \citep[SGD;][]{bottou2004} and ADAptive Moment estimation \citep[ADAM;][]{Kingma2014}, each with different variations of internal parameters such as learning rates and learning rate decay. We found that an ADAM optimiser with an initial learning rate of 0.001 and a time-based learning rate decay worked best for our application.

The neural networks we have tested are shown in Table~\ref{tab:NNets}. These networks represent different regularisation methods applied to a typical neural network. Due to the nature of regularisation, networks with regularisation tend to be bigger, with more nodes or layers.

\section{Table of SMG counterparts in GOODS-N}
\label{app:GDN_SMG}

Here we provide a table of the predicted SMG counterparts (defined as having an average prediction score over 0.5) in the GOODS-N field. Alongside the ID numbers and coordinates for both the SCUBA-2 sources and the multiwavelength sources from GOODS-N, we also provide the prediction score value and standard deviation across the 100 ensemble neural networks. The prediction score and uncertainty values help us understand the degree to which we are confident that these sources are SMGs or non-SMGs. The final columns indicate whether or not the SMG has a radio or 24\,$\mu$ counterpart.


\begin{table*}
	\centering
    \label{tab:predGDNs}
    \caption{All identified SMG counterparts in GOODS-N. Column (1) gives the names of the single-dish submm sources detected in the S2CLS catalogue \citep{geach2016}, and similarly columns (2) and (3) gives their positions (i.e.~the position of the brightest pixel within the SCUBA-2 beam). Columns (4) and (5) give the positions of the optical/NIR counterparts (taken from \citealt{hsu2019}) found in this paper using our fully trained NN, and column (6) gives the angular offset between the optical/NIR counterparts and the single-dish submm positions. In column (7) we give the prediction score for each counterpart, calculated as the mean value of the output of 100 NNs trained in parallel; numbers close to 0 are not likely SMGs, and numbers close to 1 are likely SMGs. Our threshold for positive identifications was 0.5. In columns (8) and (9) we state whether or not the single-dish submm source has a corresponding radio galaxy match or 24\,$\mu$m galaxy match, respectively (Y for yes, N for no); see Section \ref{sec:radio24} for details.}
	\begin{tabularx}{\textwidth}{cYYcYcccc}
	\toprule \noalign {\smallskip}
    S2CLS name          & RA           & Dec         & RA          & Dec        & \makecell{Angular \\ separation} & \makecell{Prediction \\ score} & \makecell{Radio \\ match}& \makecell{24\,$\mu$m \\ match}   \\
                        & \multicolumn{2}{c}{(submm)}& \multicolumn{2}{c}{(optical/NIR)}&        &        &        &        \\
                        & \multicolumn{2}{c}{[J2000]}& \multicolumn{2}{c}{[J2000]}&  [arcsec] &        &        &        \\
    (1)                 & (2)          & (3)         & (4)         & (5)        & (6)      & (7)                       & (8)   & (9)   \\
    \midrule
    S2CLSJ123730+621258 & 189.37803931 & 62.21626211 & 189.3781731 & 62.2162900 & 0.245940 & 0.747 $\pm$ 0.156 & Y     & Y     \\
    S2CLSJ123711+621330 & 189.29820699 & 62.22521367 & 189.2972740 & 62.2252994 & 1.595313 & 0.763 $\pm$ 0.158 & Y     & Y     \\
    S2CLSJ123633+621408 & 189.13961496 & 62.23575765 &             &            &          &                   & Y     & Y     \\
    S2CLSJ123707+621406 & 189.28153378 & 62.23522097 & 189.2799737 & 62.2355744 & 2.909284 & 0.827 $\pm$ 0.072 & Y     & Y     \\
    S2CLSJ123701+621146 & 189.25645425 & 62.19633929 & 189.2565799 & 62.1961887 & 0.581735 & 0.997 $\pm$ 0.002 & Y     & Y     \\
    S2CLSJ123550+621041 & 188.96004402 & 62.17830355 &             &            &          &                   & Y     & N     \\
    S2CLSJ123618+621550 & 189.07626370 & 62.26403584 & 189.0764287 & 62.2640793 & 0.317649 & 0.956 $\pm$ 0.026 & Y     & Y     \\
    S2CLSJ123632+621712 & 189.13349000 & 62.28686472 & 189.1329583 & 62.2873442 & 1.942135 & 0.990 $\pm$ 0.007 & Y     & N     \\
    S2CLSJ123741+621220 & 189.42206933 & 62.20565233 & 189.4215231 & 62.2057496 & 0.981533 & 0.981 $\pm$ 0.017 & Y     & Y     \\
    S2CLSJ123741+621220 & 189.42206933 & 62.20565233 & 189.4235441 & 62.2065473 & 4.063188 & 0.501 $\pm$ 0.190 & Y     & Y     \\
    S2CLSJ123645+621448 & 189.19088280 & 62.24689230 & 189.1910092 & 62.2463782 & 1.862851 & 0.520 $\pm$ 0.123 & Y     & Y     \\
    S2CLSJ123645+621448 & 189.19088280 & 62.24689230 & 189.1919457 & 62.2469104 & 1.783021 & 0.861 $\pm$ 0.086 & Y     & Y     \\
    S2CLSJ123627+621214 & 189.11349908 & 62.20407169 &             &            &          &                   & Y     & Y     \\
    S2CLSJ123652+621226 & 189.21833682 & 62.20745269 &             &            &          &                   & Y     & Y     \\
    S2CLSJ123627+620604 & 189.11389111 & 62.10129438 &             &            &          &                   & Y     & Y     \\
    S2CLSJ123622+621616 & 189.09413713 & 62.27127656 &             &            &          &                   & Y     & Y     \\
    S2CLSJ123645+621936 & 189.18838880 & 62.32689153 &             &            &          &                   & Y     & N     \\
    S2CLSJ123616+621514 & 189.06796115 & 62.25402645 & 189.0659326 & 62.2543111 & 3.550903 & 0.805 $\pm$ 0.114 & Y     & Y     \\
    S2CLSJ123616+621514 & 189.06796115 & 62.25402645 & 189.0671680 & 62.2538630 & 1.453721 & 0.981 $\pm$ 0.009 & Y     & Y     \\
    S2CLSJ123648+622104 & 189.20032761 & 62.35133888 & 189.2011334 & 62.3519266 & 2.507718 & 0.975 $\pm$ 0.029 & Y     & Y     \\
    S2CLSJ123635+621424 & 189.14795121 & 62.24020721 & 189.1483527 & 62.2400106 & 0.976819 & 0.994 $\pm$ 0.006 & Y     & Y     \\
    S2CLSJ123711+621208 & 189.29934611 & 62.20243532 &             &            &          &                   & Y     & Y     \\
    S2CLSJ123658+620930 & 189.24452444 & 62.15856340 &             &            &          &                   & Y     & Y     \\
    S2CLSJ123722+620538 & 189.34420182 & 62.09407222 & 189.3458748 & 62.0943551 & 2.997083 & 0.902 $\pm$ 0.071 & Y     & N     \\
    S2CLSJ123634+621922 & 189.14533585 & 62.32298343 & 189.1453858 & 62.3232032 & 0.795569 & 0.971 $\pm$ 0.014 & Y     & Y     \\
    S2CLSJ123621+621710 & 189.08809692 & 62.28627058 & 189.0873019 & 62.2860258 & 1.596289 & 0.897 $\pm$ 0.044 & Y     & Y     \\
    S2CLSJ123621+621710 & 189.08809692 & 62.28627058 & 189.0887089 & 62.2856445 & 2.475836 & 0.989 $\pm$ 0.006 & Y     & Y     \\
    S2CLSJ123554+621338 & 188.97510736 & 62.22722106 & 188.9744380 & 62.2270685 & 1.249961 & 0.587 $\pm$ 0.161 & Y     & Y     \\
    S2CLSJ123554+621338 & 188.97510736 & 62.22722106 & 188.9719555 & 62.2271217 & 5.299264 & 0.913 $\pm$ 0.041 & Y     & Y     \\
    S2CLSJ123656+621206 & 189.23501651 & 62.20189742 &             &            &          &                   & N     & N     \\
    S2CLSJ123712+621034 & 189.30404641 & 62.17632176 &             &            &          &                   & Y     & N     \\
    S2CLSJ123738+621736 & 189.41068330 & 62.29344570 & 189.4096088 & 62.2933351 & 1.842040 & 0.957 $\pm$ 0.042 & N     & N     \\
    S2CLSJ123616+620700 & 189.06750327 & 62.11680357 & 189.0688200 & 62.1174773 & 3.285908 & 0.995 $\pm$ 0.003 & N     & Y     \\
    S2CLSJ123713+621826 & 189.30795947 & 62.30743071 &             &            &          &                   & N     & N     \\
    S2CLSJ123720+621106 & 189.33502421 & 62.18519045 &             &            &          &                   & N     & N     \\
    S2CLSJ123609+620650 & 189.04138645 & 62.11399289 & 189.0391695 & 62.1131623 & 4.782771 & 0.883 $\pm$ 0.046 & N     & N     \\
    S2CLSJ123609+620650 & 189.04138645 & 62.11399289 & 189.0422881 & 62.1123289 & 6.179750 & 0.512 $\pm$ 0.126 & N     & N     \\
    S2CLSJ123721+620708 & 189.34073287 & 62.11907495 &             &            &          &                   & Y     & Y     \\
    S2CLSJ123719+621216 & 189.33032697 & 62.20463841 & 189.3317204 & 62.2058194 & 4.852592 & 0.897 $\pm$ 0.049 & N     & N     \\
    S2CLSJ123608+621440 & 189.03580211 & 62.24454100 &             &            &          &                   & Y     & Y     \\
    S2CLSJ123658+621448 & 189.24218366 & 62.24689692 &             &            &          &                   & N     & N     \\
    S2CLSJ123640+621834 & 189.16689251 & 62.30966141 & 189.1665178 & 62.3092654 & 1.557363 & 0.955 $\pm$ 0.022 & Y     & N     \\
    S2CLSJ123609+620804 & 189.03769316 & 62.13454338 &             &            &          &                   & N     & N     \\
    S2CLSJ123719+621022 & 189.33022087 & 62.17297181 &             &            &          &                   & N     & N     \\
    S2CLSJ123652+620920 & 189.21954459 & 62.15578616 &             &            &          &                   & Y     & Y     \\
    S2CLSJ123621+620718 & 189.08767445 & 62.12182560 &             &            &          &                   & Y     & Y     \\
    S2CLSJ123713+621154 & 189.30648385 & 62.19854263 & 189.3095530 & 62.1991452 & 5.591282 & 0.647 $\pm$ 0.167 & N     & N     \\
    S2CLSJ123611+621034 & 189.04933178 & 62.17622565 & 189.0479284 & 62.1760598 & 2.432530 & 0.984 $\pm$ 0.006 & N     & Y     \\
    S2CLSJ123636+621154 & 189.15282765 & 62.19854329 & 189.1503158 & 62.1984154 & 4.242634 & 0.838 $\pm$ 0.136 & N     & N     \\
    S2CLSJ123640+621004 & 189.16837377 & 62.16799542 &             &            &          &                   & Y     & Y     \\
    S2CLSJ123623+622002 & 189.09744489 & 62.33405751 & 189.0965119 & 62.3356147 & 5.818768 & 0.790 $\pm$ 0.064 & N     & N     \\
    \bottomrule 
    \noalign {\smallskip}
	\end{tabularx}
\end{table*}

\begin{table*}
	\centering
    \contcaption{}
	\begin{tabularx}{\textwidth}{cYYcYcccc}
    \midrule
    S2CLS name          & RA           & Dec         & RA          & Dec        & \makecell{Angular \\ separation} & \makecell{Prediction \\ score} & \makecell{Radio \\ match}& \makecell{24 $\mu$m \\ match}   \\
                        & \multicolumn{2}{c}{(submm)}& \multicolumn{2}{c}{(optical/NIR)}&        &        &        &        \\
                        & \multicolumn{2}{c}{[J2000]}& \multicolumn{2}{c}{[J2000]}&  [arcsec] &        &        &        \\
    (1)                 & (2)          & (3)         & (4)         & (5)        & (6)      & (7)                       & (8)   & (9)   \\
    \midrule
    S2CLSJ123652+620502 & 189.21719381 & 62.08411963 &             &            &          &                   & Y     & N     \\
    S2CLSJ123702+621422 & 189.26245740 & 62.23967131 & 189.2613696 & 62.2405352 & 3.605431 & 0.898 $\pm$ 0.048 & N     & Y     \\
    S2CLSJ123646+620832 & 189.19458136 & 62.14244890 & 189.1944602 & 62.1426222 & 0.656328 & 0.784 $\pm$ 0.134 & Y     & Y     \\
    S2CLSJ123653+621110 & 189.22429758 & 62.18634192 & 189.2237021 & 62.1867815 & 1.872107 & 0.970 $\pm$ 0.018 & N     & N     \\
    S2CLSJ123634+621240 & 189.14445393 & 62.21131622 & 189.1438694 & 62.2114090 & 1.036353 & 0.968 $\pm$ 0.023 & Y     & Y     \\
    S2CLSJ123716+621404 & 189.31850212 & 62.23464654 &             &            &          &                   & N     & N     \\
    S2CLSJ123652+621858 & 189.21829801 & 62.31634152 &             &            &          &                   & N     & N     \\
    S2CLSJ123728+621920 & 189.36780063 & 62.32238394 &             &            &          &                   & Y     & N     \\
    S2CLSJ123744+620754 & 189.43467160 & 62.13174521 &             &            &          &                           & Y     & N     \\
    S2CLSJ123800+621619 & 189.50369086 & 62.27218203 & 189.5011119 & 62.2725306 & 4.498276 & 0.568 $\pm$ 0.193 & N     & N     \\
    S2CLSJ123606+621238 & 189.02886946 & 62.21064219 & 189.0263968 & 62.2091349 & 6.831366 & 0.661 $\pm$ 0.115 & N     & N     \\
    S2CLSJ123635+620618 & 189.14712522 & 62.10520689 &             &            &          &                   & N     & N     \\
    S2CLSJ123739+621618 & 189.41294072 & 62.27177602 & 189.4091778 & 62.2714342 & 6.421896 & 0.995 $\pm$ 0.014 & N     & N     \\
    S2CLSJ123739+621618 & 189.41294072 & 62.27177602 & 189.4135752 & 62.2702138 & 5.723525 & 0.992 $\pm$ 0.011 & N     & N     \\
    S2CLSJ123622+621630 & 189.09411971 & 62.27516543 & 189.0957373 & 62.2750157 & 2.762264 & 0.510 $\pm$ 0.166 & Y     & Y     \\
    S2CLSJ123622+621630 & 189.09411971 & 62.27516543 & 189.0969088 & 62.2748190 & 4.834824 & 0.802 $\pm$ 0.123 & Y     & Y     \\
    S2CLSJ123622+621630 & 189.09411971 & 62.27516543 & 189.0943424 & 62.2749211 & 0.955394 & 0.996 $\pm$ 0.003 & Y     & Y     \\
    S2CLSJ123721+620510 & 189.34061211 & 62.08629742 &             &            &          &                   & N     & N     \\
    S2CLSJ123634+620940 & 189.14459384 & 62.16131632 & 189.1436735 & 62.1617147 & 2.109676 & 0.982 $\pm$ 0.016 & N     & N     \\
    S2CLSJ123657+621516 & 189.23860700 & 62.25467499 &             &            &          &                   & N     & N     \\
    S2CLSJ123544+621437 & 188.93440818 & 62.24380752 &             &            &          &                   & N     & N     \\
    S2CLSJ123703+620800 & 189.26590603 & 62.13355943 & 189.2670790 & 62.1320418 & 5.809055 & 0.608 $\pm$ 0.165 & N     & Y     \\
    S2CLSJ123716+620804 & 189.32058381 & 62.13464523 & 189.3225633 & 62.1346003 & 3.334661 & 0.984 $\pm$ 0.007 & Y     & Y     \\
    S2CLSJ123605+620840 & 189.02336193 & 62.14452287 &             &            &          &                   & N     & N     \\
    S2CLSJ123637+620854 & 189.15533198 & 62.14854469 & 189.1542752 & 62.1478960 & 2.934708 & 0.854 $\pm$ 0.051 & Y     & Y     \\
    \bottomrule 
    \noalign {\smallskip}
	\end{tabularx}
\end{table*}


\bsp	
\label{lastpage}
\end{document}